\documentclass[9pt,twocolumn,twoside]{osajnl}

\journal{ao} %

\setboolean{shortarticle}{false} 

\title{Demonstration of resolving power $\lambda/\Delta\lambda > 10,000$ for a space-based x-ray transmission grating spectrometer}

\author[1,*]{Ralf K. Heilmann}
\author[2]{Jeffery Kolodziejczak}
\author[3]{Alexander R. Bruccoleri}
\author[2]{Jessica A. Gaskin}
\author[1]{Mark L. Schattenburg}

\affil[1]{Space Nanotechnology Laboratory, MIT Kavli Institute for Astrophysics and Space Research, Massachusetts Institute of Technology, Cambridge, MA 02139}
\affil[2]{NASA Marshall Space Flight Center, Huntsville, AL}
\affil[3]{Izentis, LLC, Cambridge, MA}

\affil[*]{Corresponding author: ralf@space.mit.edu}

\ociscodes{(050.1950)   Diffraction gratings; (300.6560)   Spectroscopy, x-ray; (340.7480) X-rays, soft x-rays, extreme ultraviolet (EUV); (350.6090) Space optics.}

\begin{abstract}
We present measurements of the resolving power of a soft x-ray spectrometer consisting of 200 nm-period lightweight, alignment-insensitive critical-angle transmission (CAT) gratings and a lightweight slumped-glass Wolter-I focusing mirror pair. We measure and model contributions from source, mirrors, detector pixel size, and grating period variation to the natural line width spectrum of the Al and Mg K$_{\alpha_1 \alpha_2}$ doublets. Measuring up to 18$^{\rm th}$ diffraction order at characteristic Al-K wavelengths we consistently obtain small broadening due to gratings corresponding to a minimum effective grating resolving power $R_g > 10,000$ with 90\% confidence.  Upper limits are often compatible with $R_g = \infty$.  Independent fitting of different diffraction orders, as well as ensemble fitting of multiple orders at multiple wavelengths, gives compatible results.  Our data leads to  uncertainties for the Al-K$_{\alpha}$ doublet line width and line separation parameters 2-3 times smaller than values found in the literature.  Data from three different gratings are mutually compatible.  This demonstrates that CAT gratings perform in excess of the requirements for the Arcus Explorer mission and are suitable for next-generation space-based x-ray spectrometer designs with resolving power 5-10 times higher than the transmission grating spectrometer on the Chandra X-ray Observatory.
\end{abstract}

\begin{document}

\maketitle

\section{Introduction}

The soft x-ray band (roughly between 0.2 and a few keV in energy) contains many atomic resonances.  Spectra in this band offer a wealth of diagnostics about the composition, density, and temperature of x-ray emitting or absorbing objects.  In astronomy, important lines of highly ionized carbon, nitrogen, oxygen, neon and iron can be found in the wavelength range between 1 and 5 nm. Emission and absorption line spectroscopy of celestial objects and structures in this band have the potential to provide essential information for the study of large scale structure formation (galaxy clusters), feedback from supermassive black holes, hot gas in the cosmic web, and stellar evolution - information which is often not available at other wavelengths \cite{Arcus,Lynx}.

Soft x rays are readily absorbed by small amounts of matter, which makes it difficult to build efficient transmitting optical elements, such as lenses or transmission gratings.  Obviously, absorption by air requires us to study the x-ray universe from satellites above the earth's atmosphere.

Spectroscopic information can be obtained using energy dispersive instruments, such as microcalorimeters \cite{Irwin}, or grating spectrometers, which are wavelength dispersive.  The energy resolution of microcalorimeters is typically on the order of a few eV (but can be sub-eV) \cite{Bandler}, which gives $E/\Delta E ~\sim 200 - 1000$ for soft x rays.  Similar resolving power $\lambda/\Delta\lambda$ can be obtained from existing, but aging instruments on board of the Chandra (High Energy Transmission Grating Spectrometer HETG) \cite{cxc} and XMM-Newton (Reflection Grating Spectrometer RGS) \cite{RGS} x-ray observatories, both of which were launched in 1999.  Their effective areas are rather small, in the range of a few tens to $\sim 100$ cm$^2$, resulting in very long observation times up to megaseconds (over one week for a single object).  For many of the above science questions $\lambda/\Delta\lambda > 2500$ is required, and $\lambda/\Delta\lambda > 5000$ is desired. 

High-resolution soft x-ray spectroscopy has been demonstrated with laboratory sources and double-crystal spectrometers \cite{schweppe}.  In the last two decades much development has taken place at electron beam ion traps \cite{Beier1997,Widmann,Yang,Beier2018} and synchrotron sources, the latter mostly focusing  on resonant inelastic x-ray scattering \cite{Ghiri,Strocov,Harada,Warwick,Qiao,Sala}. Dispersing elements are mostly crystals, plane, or variable-line-spacing reflection gratings. The spectrometers often achieve $\lambda/\Delta\lambda$ on the order of a few thousand to $\sim 10000$.  Many of these spectrometers depend on strong sources, precise and adjustable alignment, and multiple movable elements to achieve a broad bandpass.  These designs would be difficult to implement in space, where movable elements and mass should be minimized.

Space-based x-ray grating spectrometers (XGS) are typically designed with an array of objective gratings just downstream of the focusing telescope mirrors (the ``lens'' - usually a set of concentric Wolter-I grazing-incidence mirrors).  Due to the sparseness of celestial x rays, mirrors should extend over a significant aperture on the order of 1 m$^2$, and gratings should cover a large part of or the whole mirror aperture. In the in-plane transmission geometry, where the grating vector connecting two grating bars lies in the plane of incidence, the gratings diffract photons incident at angle $\alpha$ relative to the grating normal into diffraction orders $m$ at angles $\beta_m$ according to the grating equation
\begin{equation}
\frac{m \lambda}{p} = \sin \alpha - \sin \beta_m ,
\label{ge}
\end{equation}
\noindent
where $m = 0, \pm 1, \pm 2,...$, $\lambda$ is the x-ray wavelength, and $p$ is the grating period (see Fig.~\ref{fig:CAT}).  The gratings are arrayed on the surface of a Rowland torus \cite{SPIE2010,Moritz17}, such that the $m^{\rm th}$ diffraction order from each grating comes to a common focus on the surface of a detector with fine spatial resolution, typically an x-ray CCD.  For a broad spectrum, different orders from different wavelengths can overlap spatially.  The resulting limited free spectral range $\Delta \lambda = \lambda /m$ can be overcome if the energy resolution of the detector is better than the corresponding photon energy difference $\Delta E$, divided by $m$.\cite{AO}

The resolving power of an XGS can be defined as
\begin{equation}
R_{XGS} = \lambda/\Delta\lambda ,
\end{equation}
\noindent
where $\Delta\lambda$ is the smallest wavelength difference that can be resolved at wavelength $\lambda$.
To first order, $R_{XGS}$ is given by the distance of the $m^{\rm th}$ order spot on the CCD from the zeroth order divided by the width of the telescope mirror point-spread function (PSF) in the dispersion direction.  It is therefore advantageous to use high diffraction orders and small grating period and to have a narrow PSF.  High diffraction orders are only useful if a large percentage of incident photons lands in these orders.  This has been achieved via blazing with sawtooth groove profiles for grazing incidence reflection gratings \cite{RGS,Chi,Seely,Randy1,Randy2}.  However, the reflection geometry is very sensitive to misalignments and grating non-flatness, and grazing incidence requires many-cm long substrates with larger mass than $\mu$m-thin transmission gratings.

Critical-angle transmission (CAT) gratings combine the advantages of the transmission geometry (alignment insensitivity, low mass) with efficient utilization of high diffraction orders (blazing) \cite{OE,AO}.  As shown in Fig.~\ref{fig:CAT}, this is accomplished by tilting freestanding ultra-high aspect ratio grating bars by a small angle $\alpha$, which is less than the critical angle for total external reflection, relative to the incident x rays.  Diffraction orders near the direction of specular reflection from the grating bar sidewalls have enhanced diffraction efficiency.  Thin grating bars ($b < p/3$) and lack of a support membrane minimize absorption.
We have recently fabricated 200 nm-period, 4 $\mu$m deep silicon CAT gratings up to $32 \times 32$ mm$^2$ in size, with blazed diffraction efficiency $> 30$\% at $\lambda \sim 2.5$ nm and $> 20$\% for 1.5 nm $< \lambda < 5$ nm,\cite{SPIE2017} compared to $\sim 1-5$\% for HETG gratings.\cite{cxc}

At $\lambda = 1.5$ nm the critical angle for silicon is $\sim 2$ degrees.  If we set $\alpha = 2$ degrees then we expect to blaze orders near $\alpha + \beta_m = 4$ degrees, i.e., $9^{\rm th}$ and $10^{\rm th}$ order.  For a telescope with a PSF of 1 arcsec full width half max (FWHM) $f_{PSF}$, we expect $R_{XGS} \approx (\alpha + \beta_m)/f_{PSF} = 4^{\circ}/1" = 14400$ (neglecting that the gratings are slightly closer to the focus than the mirrors).  However, XGS optical designs are neither free from aberrations, nor will a real XGS follow its design perfectly.  We have undertaken numerous ray-trace studies of transmission XGS designs to understand the limits of performance, alignment tolerances, and other imperfections \cite{SPIE2010,Davis,Bautz,Moritz16,Moritz17,MoritzLynx18}, and came to the conclusion that instruments with $R_{XGS} \sim 10000$ and effective area $> 1000$ cm$^2$ should be feasible in the very near future.

\begin{figure}
   \begin{center}
   \includegraphics[width=2.9in]{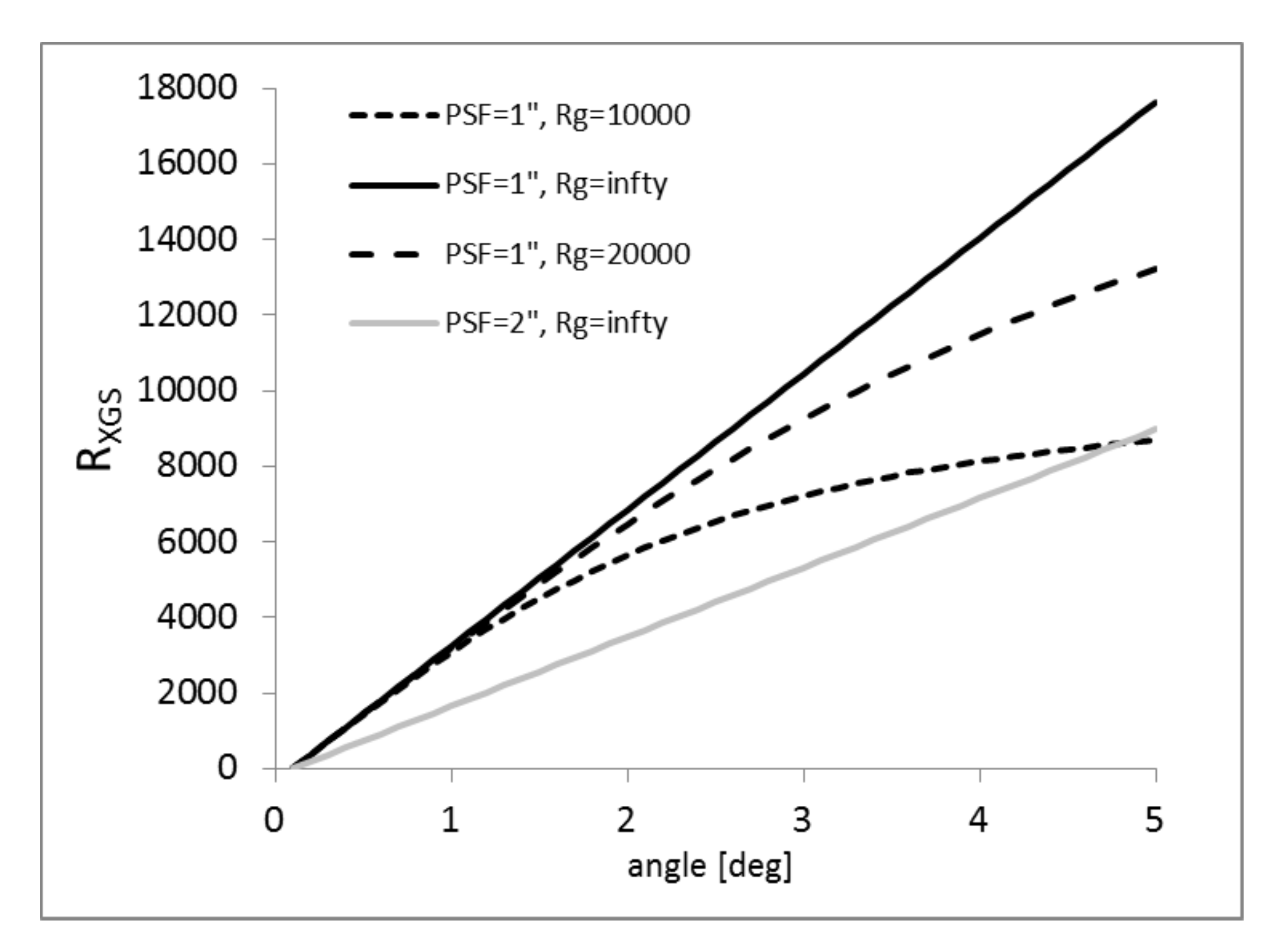}
   \end{center}
   \caption[holder] 
   { \label{fig:RXGS} 
\small Simple model of resolving power as a function of diffraction angle for an objective grating spectrometer with $f_{PSF} = 1$ arcsec (black lines) and 2 arcsec (grey line).  Solid lines represent no broadening due to the grating (``$R_g = \infty$"), while dashed lines show the impact of finite effective grating resolving power $R_g$.}
   \end{figure}

XGS resolving power can be compromised by grating imperfections, such as variations in the grating period, described by some period distribution $\{p\}$ with FWHM $\Delta p$, for example.  Eq.~\ref{ge} shows that if $\Delta p \ne 0$ then there will be a distribution of diffraction angles $\{\beta_m\}$ with FWHM $\Delta \beta_m$ and a broadening of the $m^{\rm th}$ order diffraction peak proportional to $m$.  Neglecting aberrations, the observed peak broadening is then a convolution between the PSF and the $\beta_m$ distribution function.  Since $\Delta \beta_m$ scales with $m$ it can become the dominating source of broadening in higher orders and limit resolving power to a value less than $(\alpha + \beta_m)/f_{PSF}$.  If we assume $\{p\}$ to be Gaussian, then $R_{XGS}$ can never be greater than $p/\Delta p$.  In our analysis we simply model grating imperfections as a Gaussian period distribution and call $R_g = p/\Delta p$ the {\em effective resolving power} of the grating (see Fig.~\ref{fig:RXGS}), which is different from the traditional definition of resolving power or resolvance of a grating, $mN$, where $N$ is the number of illuminated grating lines.
The present work was undertaken to study broadening of spectral features due to potential CAT grating imperfections that could limit resolving power in an XGS.  

   \begin{figure}
   \begin{center}
   \includegraphics[width=3.5in]{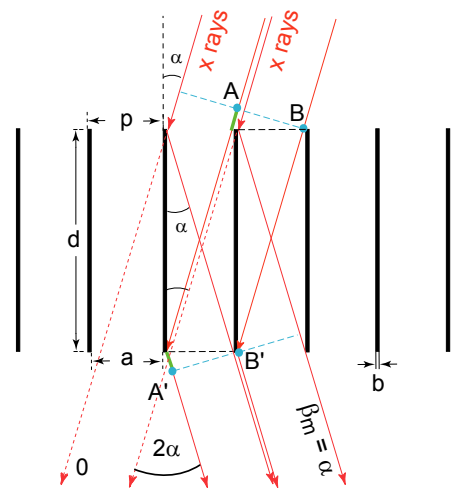}
   \end{center}
   \caption[holder] 
   { \label{fig:CAT} 
\small Schematic cross section through a CAT grating of period $p$.  The $m^{\rm th}$ diffraction order occurs at an angle $\beta_m$ where the path length difference between AA' and BB' is $m\lambda$.  Shown is the case where $\beta_m$ coincides with the direction of specular reflection from the grating bar sidewalls ($|\beta_m| = |\alpha|$), i.e., blazing in the $m^{\rm th}$order.}
   \label{fig:CAT}
   \end{figure} 

To mimic an XGS we need a soft x-ray source with a narrow spectral line, a focusing optic with narrow PSF that can fully illuminate a grating of reasonable size, and a detector with high spatial resolution.  At the time of this work we were not aware of any synchrotron endstations that could have provided us with an expanded, collimated beam and a 10-m long vacuum chamber and the necessary manipulators.  For traditional laboratory soft x-ray sources the narrowest lines are provided by the Al and Mg K$_{\alpha_1 \alpha_2}$ doublets at $\lambda = 0.834$ and 0.989 nm, respectively, with $E/\Gamma \sim 3500$, where $\Gamma$ is the FWHM of each of the K$_\alpha$ lines and $E$ is the photon energy.  For silicon the critical angles for these wavelengths are only $\sim 1.1$ and $\sim 1.35$ degrees.  In order to obtain higher diffraction efficiencies at the highest orders possible we coated some CAT gratings with a thin layer of platinum, effectively increasing the critical angle. 

In the following, we first describe our experimental setup, then our measurements and models, and finally the results of fitting our models to the data.  We then discuss our results and summarize our conclusions.

\section{Experimental Setup}
\label{sec:setup}

\begin{figure}
 \begin{center}
   \includegraphics[width=3.5in]{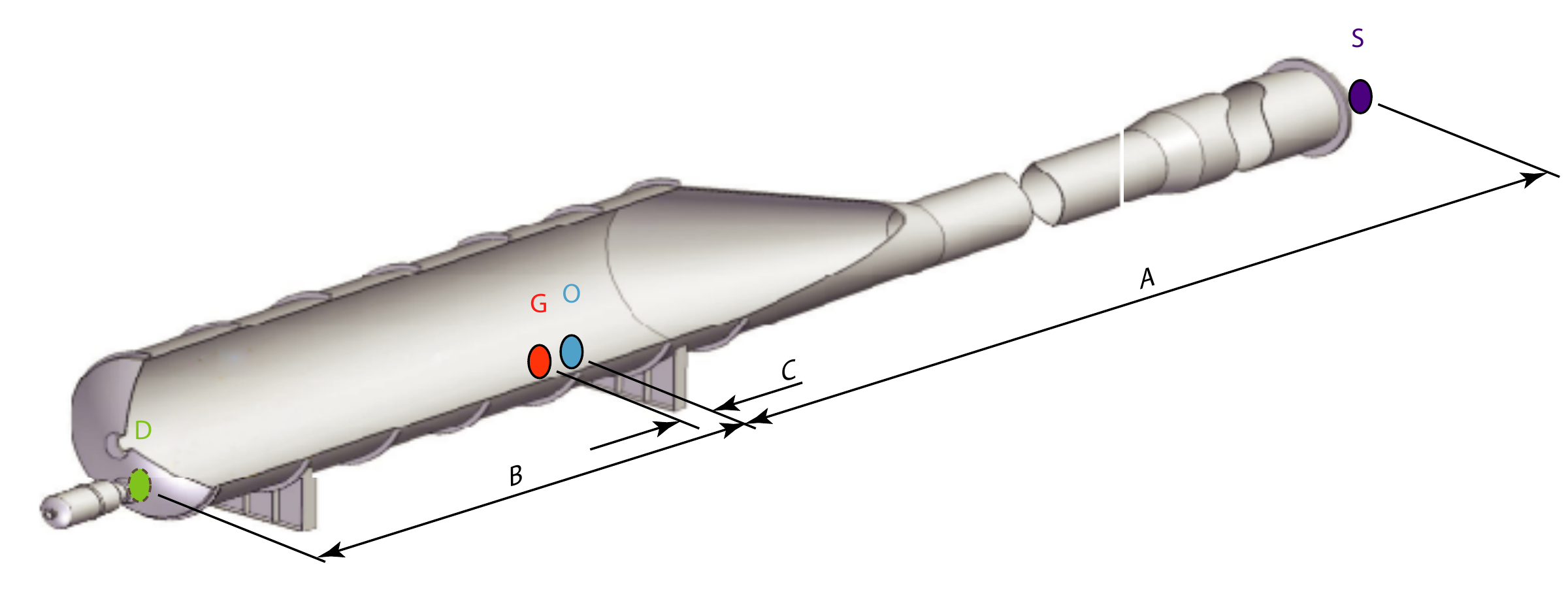}
 \end{center}
\caption{\small Overall configuration of the SLTF for this test, depicting the locations of S-source, O-optic node, G-grating, and D-detector. The source-to-optic-node distance (A) is $92.17\,{\rm m}$. The detector-to-optic-node distance (B) is $9.25\,{\rm m}$. The grating-to-optic-node distance (C) is $0.50\,{\rm m}$. For reference, the source is at the east end and the detector is at the west end of the facility.}
 \label{slffig}
\end{figure}

The measurements were performed at the NASA Marshall Space Flight Center Stray Light Test Facility (SLTF).  It consists of a 92 m long, 1.22 m-diameter vacuum guide tube that opens into a 12.19 m-long, 3.05 m-diameter vacuum chamber (see Fig.~\ref{slffig}).  The far end of the guide tube connects to a Manson electron impact x-ray source.  Both Al and Mg anodes were used in this work.  The source is equipped with 100 $\mu$m and 150 $\mu$m vertical slits to reduce the horizontal width of a 0.5 mm source spot, effectively improving the source size from 1.12 arcsec to 0.22 and 0.34 arcsec, respectively.

   \begin{figure}
   \begin{center}
   \includegraphics[width=2.2in]{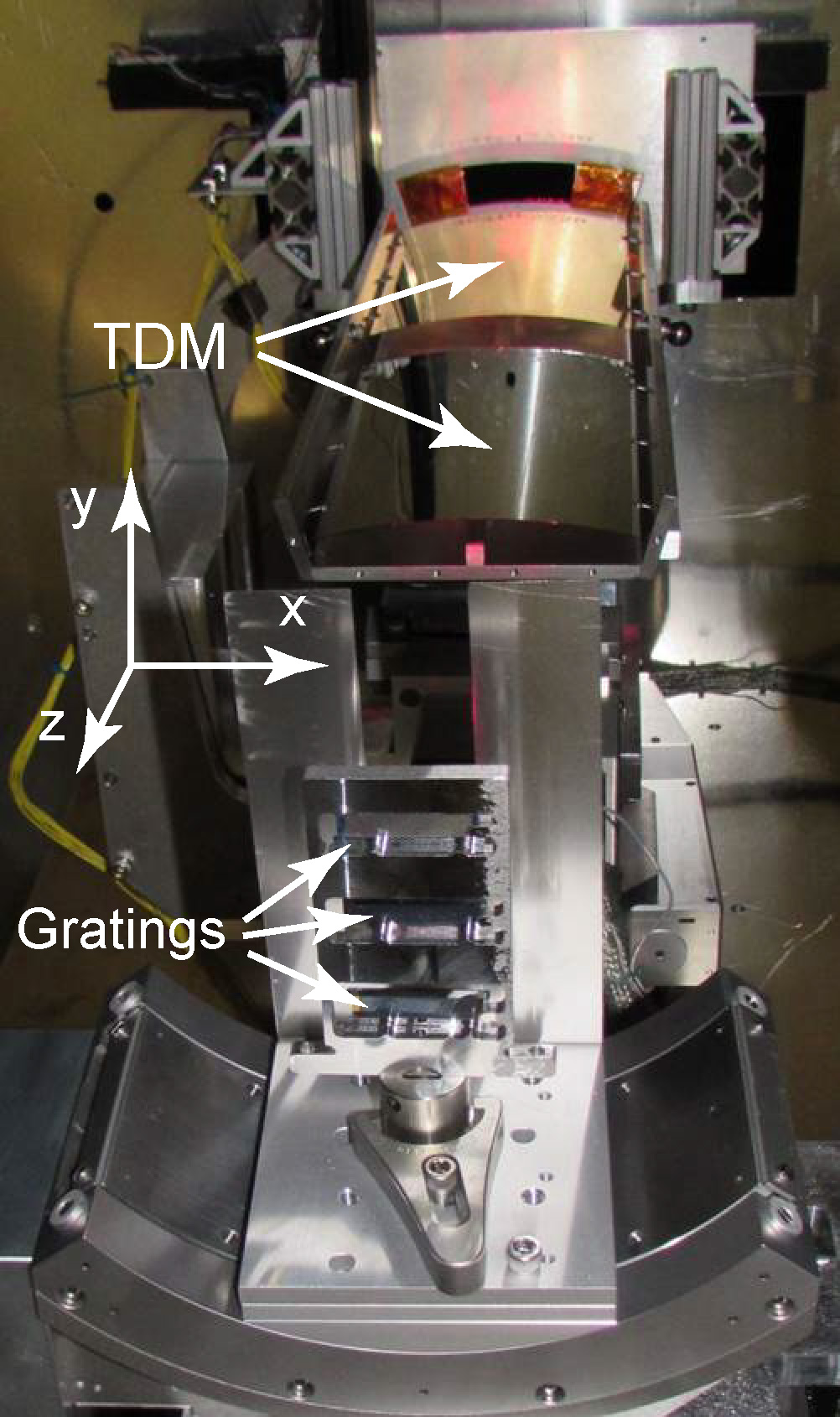}
   \end{center}
   \caption[holder] 
   { \label{fig:setup} 
\small Picture of the experimental setup.  The mount with the three gratings is seen in the foreground, followed by the 30 mm grating mask, the Ir and Au coated TDM mirrors, and the optics mask.}
   \end{figure} 

The grating spectrometer is part of an imaging system.  In astronomical applications an ideal point source at infinity is imaged to a small spot in the focal plane of a focusing optic.  The image of the point source is broadened due to the finite angular width of the optic PSF.  In order to separate spectral features with a small $\Delta \lambda$, a sufficiently small PSF is required.  In this work we used an 8.4 m focal length Technology Development Module (TDM) \cite{TDM} manufactured by the Next Generation X-ray Optics group at the NASA Goddard Space Flight Center.  The TDM consisted of two 0.4 mm-thick Wolter-I slumped glass segments: a gold-coated parabolic (P) segment, followed by an iridium-coated hyperbolic (H) segment.  The radius at the node of the optic - the center between the P and the H segments - was 245 mm, the azimuthal extent 30 degrees, and each glass segment was about 200 mm long along the optical axis or $z$ direction (see Fig.~\ref{fig:setup}).  The half-power diameter (HPD) of the 2D PSF of a single mirror-pair TDM is typically about 8 arcsec, and the coating roughness about 0.5 nm.  Due to the finite source distance the best focus of the source is obtained 9.25 m from the optic node.  The optic was mounted on a stage stack for pitch (rotation around $x$ or the horizontal axis) and yaw (rotation around $y$ or the vertical axis) alignment.  Upstream of the optic was a large plate to block direct illumination of the focal plane by the source, and a selectable aperture plate for the TDM.

\begin{figure}
   \begin{center}
   \includegraphics[width=3.5in]{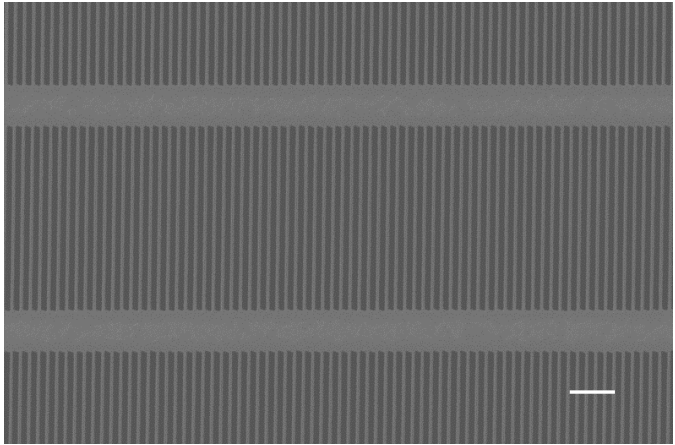}
   \end{center}
   \caption[holder] 
   { \label{fig:SEM} 
\small Top down scanning electron micrograph of grating X7, showing the 200-nm period CAT grating bars and the 5 $\mu$m period L1 cross support mesh. The scale bar is 1 $\mu$m long.}
   \end{figure}

\begin{figure}
   \begin{center}
   \includegraphics[width=3.5in]{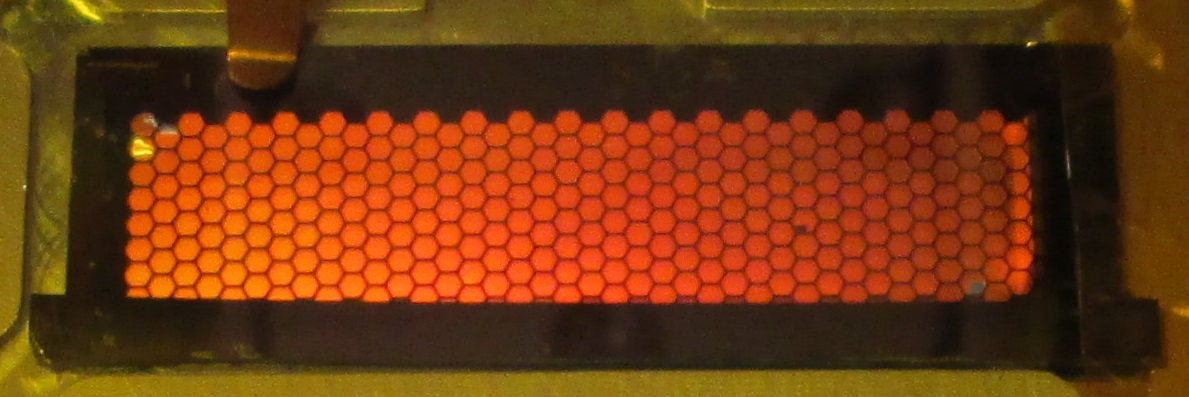}
   \end{center}
   \caption[holder] 
   { \label{fig:X7} 
\small Picture of grating X7.  The device layer is partially transparent, making the back-lit hexagonal L2 support mesh visible. The L1 and L2 support structures combined occupy about 34\% of the grating area.}
   \end{figure}

Three different CAT gratings were used in this work.  They were all fabricated from silicon-on-insulator (SOI) wafers.  Freestanding CAT grating bars with 200 nm period were etched out of the nominally 4 $\mu$m-thick SOI device layer at the same time as a 5 $\mu$m-period Level 1 (L1) cross-support mesh (see Fig.~\ref{fig:SEM}), using a combination of deep reactive-ion etching (DRIE) and wet etching in potassium hydroxide solution.  A coarse Level 2 (L2) hexagonal support mesh ($\sim 1$ mm pitch) was etched out of the 0.5 mm-thick SOI handle layer.  The buried oxide layer separating device and handle layers has been removed from the open areas between the etched structures.  Details of the fabrication process can be found in Refs.~\cite{JVST16,JVST13,AlexSPIE2013,JVST12,Ahn07}.  The usable grating area was 32 mm long and between 5.7 and 7.5 mm wide, depending on the grating.  Gratings X1 and X4 were nominally coated with 2 nm of aluminum oxide and 7 nm of platinum using atomic layer deposition.  Grating X7 was left uncoated (see Fig.~\ref{fig:X7}).


The gratings were mounted in a vertical stack in a grating holder 50 cm downstream of the TDM node, with their long axes in the horizontal direction (see Fig.~\ref{fig:setup}).  The holder was mounted on a stage stack consisting of a linear x-translation stage at the bottom that carried a yaw (rotation around vertical axis) stage, followed by y-translation and roll (rotation around optical axis) stages.  The whole stage stack was tilted in pitch by $\sim 1.4$ deg to achieve close to normal x-ray incidence on the gratings.  Just upstream of the gratings we placed an aperture mask that limited the grating illumination to 30 mm in the horizontal direction.  The converging x-ray beam incident on a grating thus had a cross section in the shape of a shallow arc of about 1.5 mm in radial extent and about 30 mm in the azimuthal direction and 242 mm radius of curvature.

The detector was a model DX436-BN-9HS CCD manufactured by Andor Corp. The imager consists of 2048x2048 pixels (13.5 $\mu$m pixel pitch) with settable clocking speeds. The array was covered by an optical blocking filter consisting of 150 nm Al on 200 nm polyimide.  Needing minimal energy resolution, we ran at the fastest clocking speed of 1 $\mu$s per pixel during the test. We operated in subarray mode for a portion of the test with a 1024 x1024 subarray configured.  The detector was mounted on a 3-axis xyz stage stack. The detector/stage system was displaced 152 mm from the center of the chamber toward the south to enable grating measurements at larger angular dispersion. The detector operating temperature was maintained at a constant -45$^o$C for the entire test.

\begin{figure}
   \begin{center}
   \includegraphics[width=3.5in]{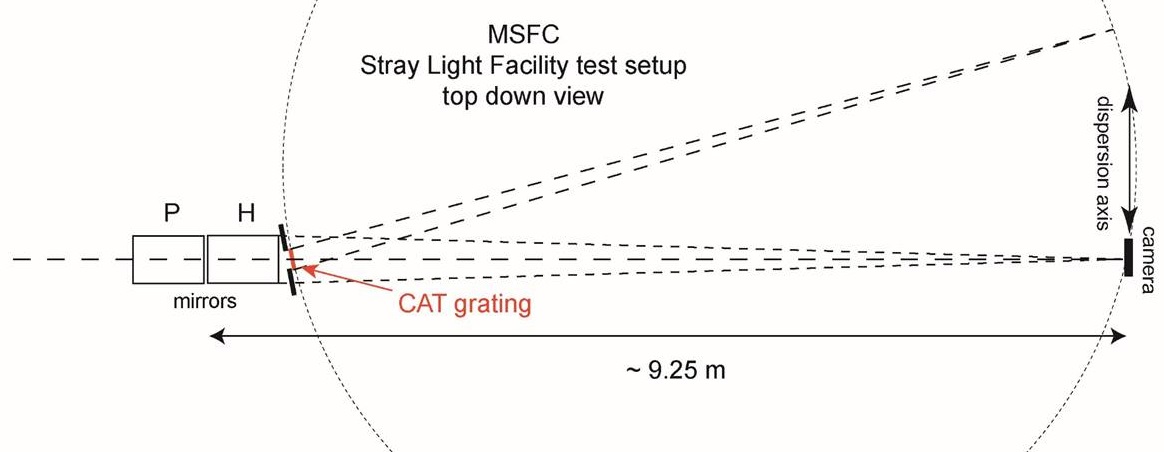}
   \end{center}
   \caption[holder] 
   { \label{fig:layout} 
\small Planview of the experimental layout (not to scale).  X rays are incident from the source from the left.  The dotted circle shows a cross-section of the Rowland torus.}
   \end{figure}

Before evacuation of the chamber we performed preliminary alignment of the optics, gratings, and masks using a HeNe laser at the source end.  The TDM was placed 245 mm above the horizontal optical axis with its reflective sides facing down.  A second laser aimed from the optics focus back towards one of the gratings was used to visualize the orientation of the L1 mesh dispersion axis.  We rolled the grating mount until this axis was vertical, thus placing the CAT grating dispersion axis close to horizontal orientation.  The gratings therefore disperse close to the direction along which the anisotropic optic PSF is expected to be at its narrowest.  A schematic of the experimental layout is shown in Fig.~\ref{fig:layout}.

\section{Measurements}
\label{meas}
We first characterized the direct (``unobstructed'') beam from the TDM, and after grating insertion the beam transmitted straight through the gratings (0$^{\rm th}$ diffracted order).  Following this we explored higher diffraction orders until hardware limitations prevented us from going further.

Dark images (with the source shutter closed) were collected periodically to monitor the dark levels produced by the camera readout electronics. The first images of the optic under Al-K illumination gave a sufficiently narrow PSF.  We did not perform any further pitch and yaw fine adjustment of the optics with x rays until the end of this study.  The best focus was found through a series of images taken at different camera positions along the optical axis.

\subsection{Combined Performance of Source, Mirror, and Detector}
\label{sourcemirror}
\begin{figure}
 \begin{center}
   \includegraphics[width=3.5in]{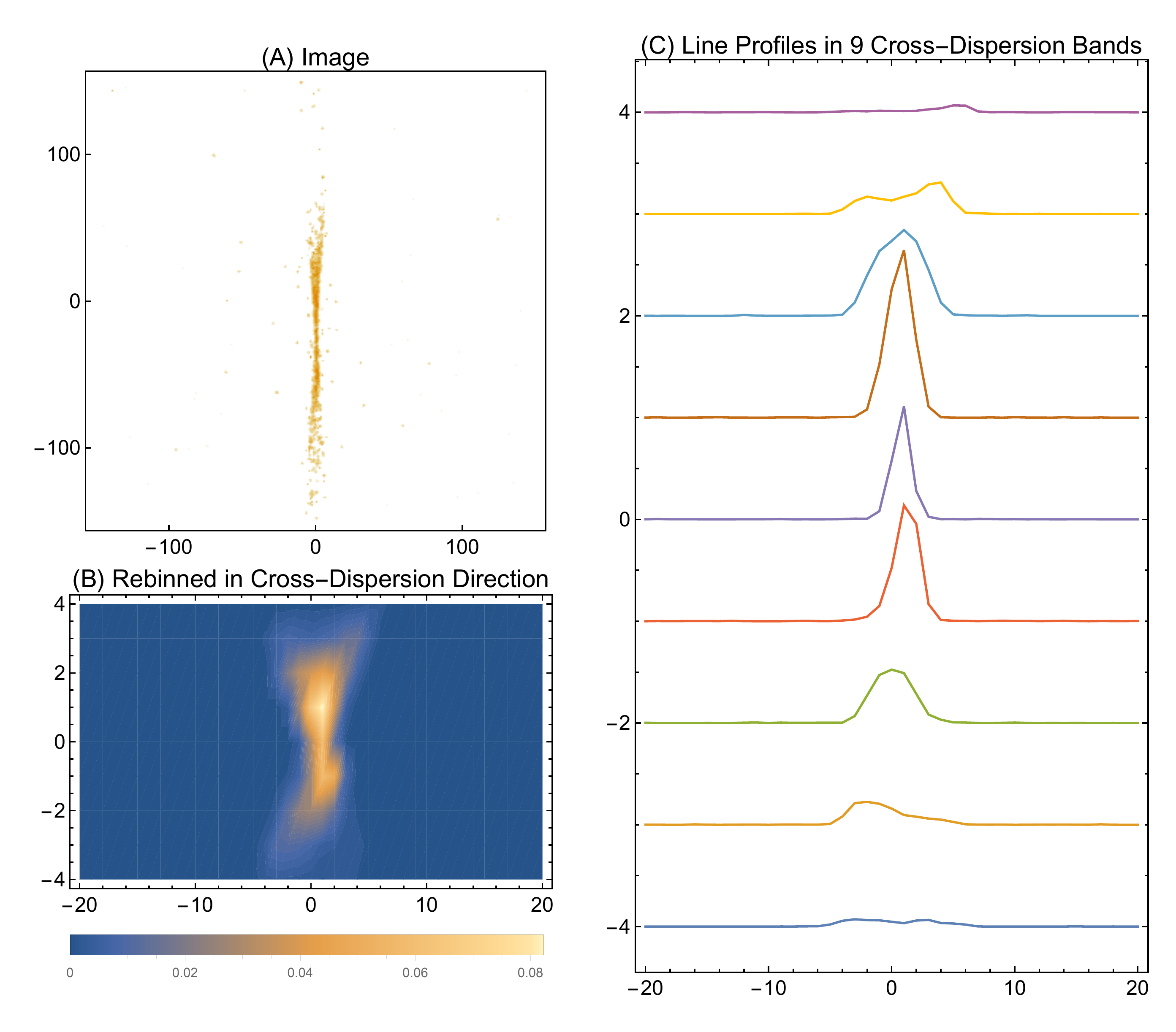}
 \end{center}
\caption{\small Conversion of unobstructed beam image to line profiles for $100\rm{\,\mu m}$ slit data. (A) Original image showing ``bow-tie" structure of mirror PSF.  (B) Image recentered and rebinned into 9 bands around the minimum FWHM. Each band is 21 pixels tall.  (C) Stacked and relatively normalized line profiles for each band. The band with the smallest FWHM does not have the highest flux in this case.}
 \label{rebin}
\end{figure}

The focal spot from the TDM (direct beam) exhibits a narrow ``hour-glass" or rotated ``bow-tie" cross section whose dispersive-direction (x) FWHM varies as a function of cross-dispersion direction coordinate (y) (see Fig.~\ref{rebin}(a)).  This so-called sub-aperture effect is a well-known feature of reflection at small angles of grazing incidence from surfaces of finite roughness.\cite{SPIE2010,MoritzLynx18,cash}  The measurement is the result of the convolution of the source size and the mirror PSF, and finite CCD pixel size.  The mirror performance is estimated from images taken under three source slit configurations. The open (no slit), 150 $\mu$m, and 100 $\mu$m configurations, using the Al anode, produce sources with FWHMs of 0.97", 0.33" and 0.22" in the dispersion direction, respectively, indicating a 0.43 mm source spot width.  The source spot extends ~1.2" in the cross-dispersion direction in all cases.  

\begin{figure}
 \begin{center}
   \includegraphics[width=3.5in]{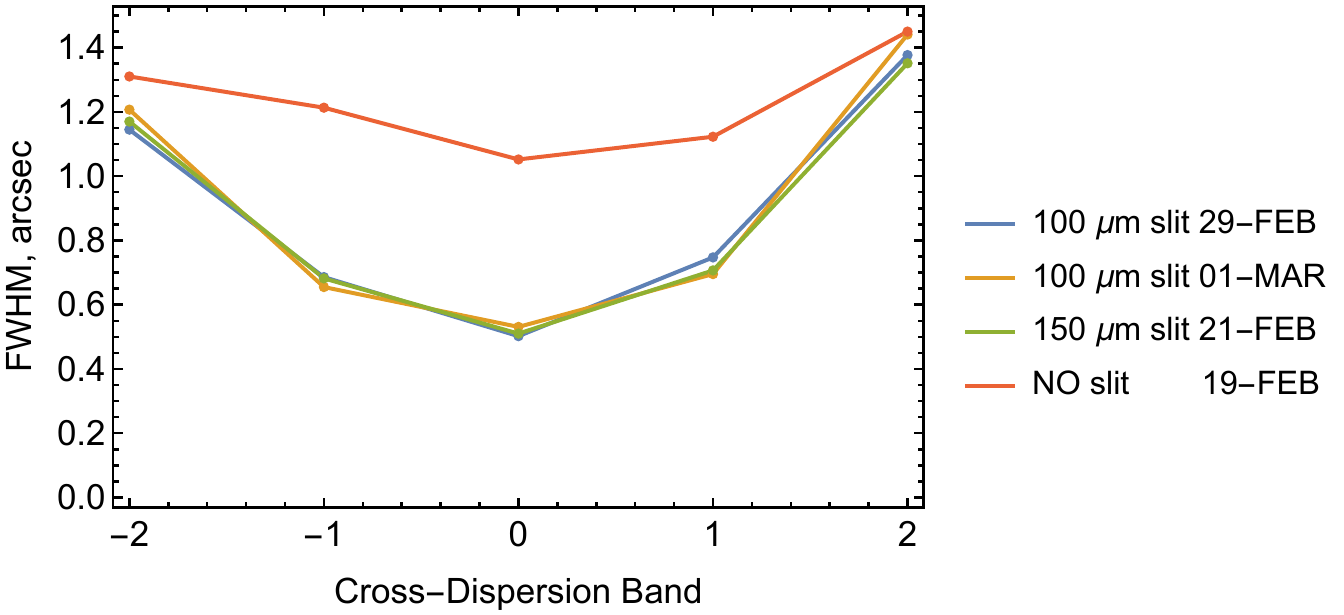}
 \end{center}
\caption{\small Measured line profile FWHM vs. CDB for unobstructed images obtained with the indicated slit configurations on indicated dates.  The effect of the 100 $\mu$m slit, in comparison to the 150 $\mu$m slit is negligible, both being $0.50^{\prime\prime}$ at minimum. The minimum FWHM for the no-slit configuration was $1.05^{\prime\prime}$. $1\sigma$ errors of these values are less than $0.05''$. }
 \label{FWHMs}
\end{figure}

Due to the irregular shape of the beam image we define a series of cross-dispersion bands (CDBs).  These CDBs are a re-binning of images into nine 21-pixel vertical bands extending from -94 to +94 pixels from the narrowest region of the bow-tie. We determine the narrowest region of the bow-tie from the FWHM in the dispersion direction based on 2.35 $\sigma$.  We define the term ``line profile" as the 1-dimensional measured distribution of detected charge in the dispersion direction integrated (or binned) over some range in cross-dispersion direction.
Fig.~\ref{rebin}(b) shows an image transformed to a binned image along with a series of line profiles in various CDBs (Fig.~\ref{rebin}(c)). Fig.~\ref{FWHMs} is a plot of the measured FWHM vs.~CDB for a representative set of images taken at different times. The central five bands are consistent and were used in our modeling.  However, the remaining bands, representing poorer regions of the mirrors, varied significantly (not shown). Their line profiles varied from single peak to double peak at different times and are not useful for measuring grating performance. We conjecture that the cause of this variation is thermal, and that regions of the optics which have larger slope errors, e.g., the ends and regions near mounting supports, are more thermally sensitive. (Temperatures in this region of the chamber typically vary by several degrees F during the course of the day.)  Fit over the five central CDBs, the rotation angle of the bow-tie relative to the CCD columns is < 0.4 deg.
As seen in Fig.~\ref{FWHMs}, the $\pm2$ CDBs do not benefit from insertion of the source slits, and the difference between the two slit configurations is negligible across all five central CDBs.  In hindsight analysis showed that both slits, combined with the 0.30" pixel width, produce nearly indistinguishable line profiles when convolved with the mirror PSF, and there was no benefit in using the narrower slits.

From Fig.~\ref{FWHMs} and the grating equation we can estimate $R_{XGS}$, neglecting any potential broadening from the gratings.  For characteristic Al $K_{\alpha}$ radiation in 18$^{\rm th}$order one obtains $R_{XGS} \sim 17200$ when using all five central CDBs.  However, $R_{XGS}$ can be increased to $\sim 24000$, for example, simply by utilizing only the three central CDBs, at the cost of losing counts and increasing statistical uncertainty. In principle the same trade-off between effective area and resolving power can be exercised post-observation in the analysis of data from a space-based x-ray spectrometer.

\begin{figure}
 \begin{center}
   \includegraphics[width=3.0in]{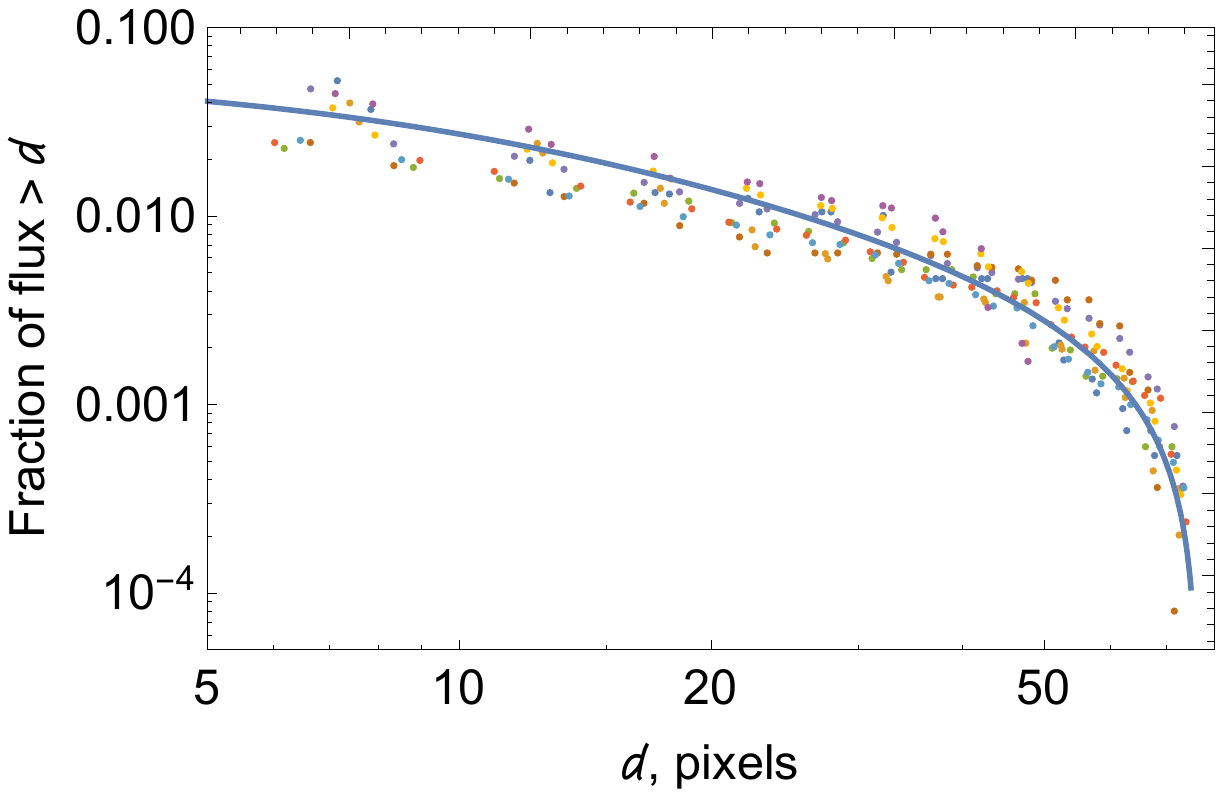}
 \end{center}
\caption{\small Scatter plot of cumulative distributions of scattering along the dispersive direction from various band/slit combinations from the sets $\{\pm2,\ \pm1,\ 0\}$ bands and $\{150\,\mu{\rm m},\ 100\,\mu{\rm m},\ {\rm open}\}$ slit.  The solid curve is a Lorentzian function with ${\rm FWHM}=20$ pixels.  The plot indicates that $\sim4\%$ of the flux is scattered in the range 5 to 75 pixels, and we estimate roughly another $\sim0.5\%$ outside of 75 pixels.}
 \label{unobstrWings}
\end{figure}

We also examined the wings of the line profiles, which we can later evaluate for any additional scattering introduced by the gratings. At grazing incidence surface roughness produces primarily in-plane scattering; the out-of-plane scattering, i.e., scattering along the grating dispersion direction, is primarily due to roughness and particulate contamination on the mirror surfaces \cite{Jeff,ODell}. Fig.~\ref{unobstrWings} indicates that the cumulative distribution function (CDF) of the scattering distribution is consistent among all the slit configurations and CDBs and that it can be modeled by a Lorentzian function. From this model we estimate that < 5\% of the flux is scattered beyond 5 pixels (or 1.5") along the dispersion direction.

\subsection{Mirror Line Response} 

Knowing the contribution of source slit width or effective spot size, and detector pixel size to the line profile, we can back out the mirror response by root difference of squares. The so-derived FWHMs and half power widths (HPW) are included in Table~\ref{bandTable}.  
The change with cross-dispersion is simply a function of the bow-tie wedge angle, due to the $7^{\circ}$ angular extent of the used aperture.  84\% of the flux is fairly evenly distributed among the 5 central CDBs, while 15\% is in the central band. The cross-dispersed flux distribution was double-peaked, so at best focus the two peaks straddled the center of the bow-tie, leaving a lower intensity at the center. (Small TDM realignment near the end of testing eliminated the double-peak, and increased the flux in band 0 by > 50\%.)

\begin{table}[h]
   \centering
{\small \begin{tabular}{lccc} \hline 
\textbf{Cross-Dispersion} & \textbf{Line} &  \textbf{Half Power}  &  \textbf{Flux Fraction} \\ 
 \textbf{Range}&  \textbf{FWHM} & \textbf{Width}  &  \textbf{in Range} \\ 
arcsec (pixels)  &  arcsec &  arcsec & $\%$\\ \hline \hline
 &&&{\vspace{-0.07in}}\\
 \text{$<3.16\,(<10.5)$} & $0.40\pm0.01$ &  $0.27\pm0.01$ &  $15$ \\
 \text{3.16--9.48 (10.5--31.5)} & $0.61\pm0.03$  &  $0.41\pm0.02$ &  $36$\\
\text{9.48--15.80 (31.5--52.5)} & $1.23\pm0.12$ &  $0.84\pm0.08$ &  $33$\\
\end{tabular}}
  \caption{\small \textbf{Mirror Line Response in 3 CDBs.}
FWHM, half-power width, and fractional flux are listed for the central 5 CDBs. The results from each of $\pm1$ and $\pm2$ CDBs are averaged, and differences are included in the errors. The half-power width along the cross-dispersion direction is $19''$, and $84\%$ of flux is within the central 5 CDBs.}
   \label{bandTable}
\end{table}

\subsection{Estimated Flux and Effective Area}

We have previously measured the x-ray flux from the Manson source with the same detector on several occasions and obtained repeatable results.  Based on facility geometry, anode voltage, slit settings, and measured count rates in the Al K$_{\alpha}$ band through the TDM we find an effective area of 0.36 cm$^2$ for the mirror pair.  This is about 90\% of the geometric aperture, consistent with 95\% reflectivity per mirror, and slightly high compared to theoretical reflectivities (92-93\%).  However, we estimate at least 5\% uncertainty for this measurement.  

\subsection{Zeroth Order Profiles}

The zeroth order line profiles are insensitive to narrow features in the source spectrum.  They were measured to investigate any non-dispersive effects the gratings may have on the line profile.  

\begin{figure}
 \begin{center}
   \includegraphics[width=3.2in]{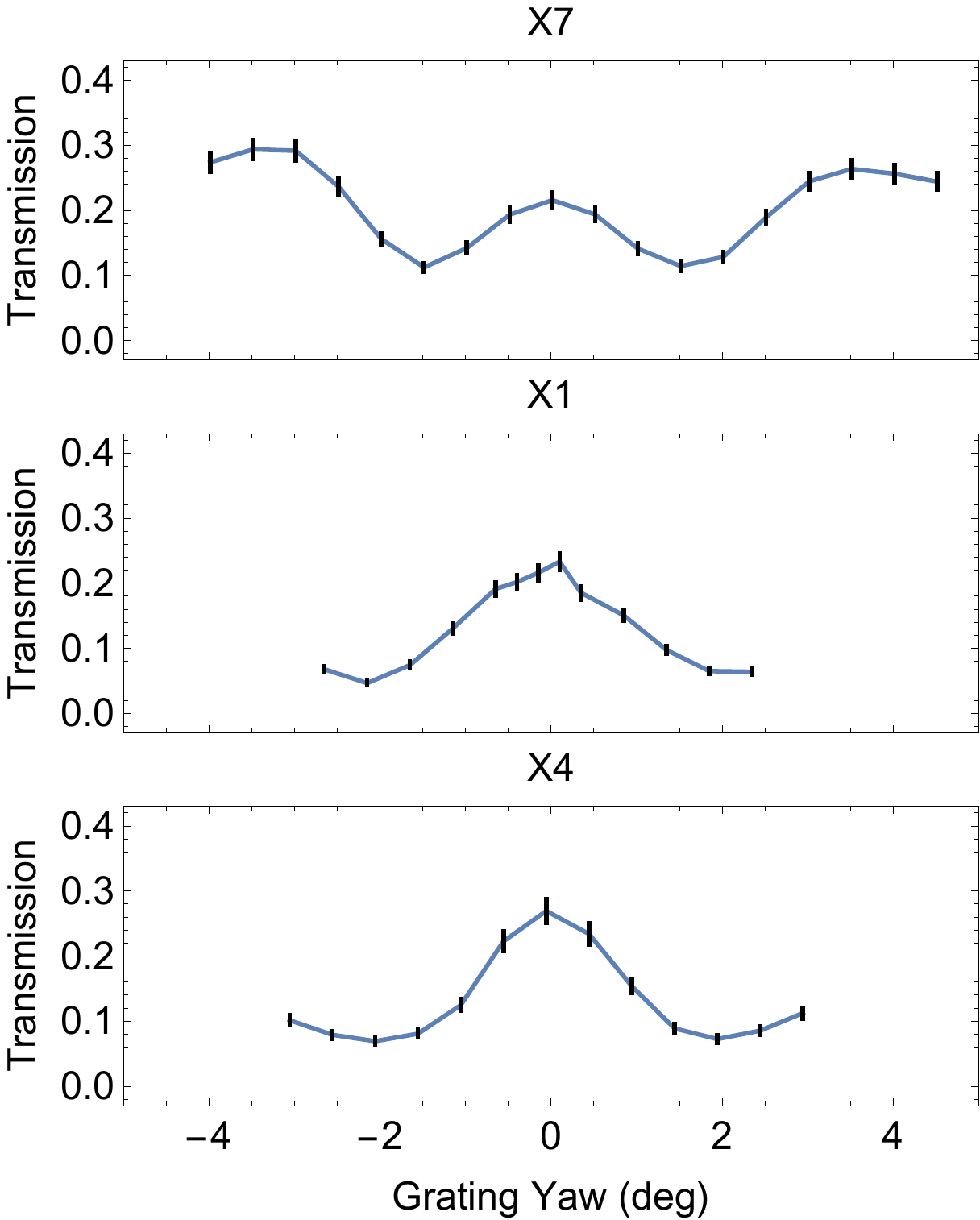}
 \end{center}
\caption{\small Measured transmitted zero-order diffraction efficiency as a function of grating yaw for the three gratings. Curves are shifted to place the central maximum at zero.  Maximum transmission values at zero yaw were: ${\rm X7:}\,0.22\pm0.01$, ${\rm X1:}\,0.23\pm0.01$, and ${\rm X4:}\,0.27\pm0.02$.}
 \label{gAlign}
\end{figure}

The vertical stage on the grating stage stack was used to move gratings in and out of the beam, and the beam was centered on the grating during illumination.  For blazed gratings the efficiency of diffracted orders is sensitive to alignment,\cite{SPIE2009,OE,AO} which can vary between gratings due to our mounting method.  We thus measured the zero-order flux as a function of yaw angle (rotation around an axis parallel to the grating bars and the grating surface, which sets the blaze angle).  For ideal CAT gratings this function is symmetrical around the angle of normal incidence.  Fig.~\ref{gAlign} shows measured zero-order efficiency for all three gratings, offset in yaw to center the curves on zero degrees.
Differences between the gratings are due to small differences in average grating bar widths and device layer thickness, as well as the Pt coating for gratings X1 and X4. 

\subsubsection{Comparison with mirror-only and among gratings}

Zeroth order line profiles were measured with each grating at its own (yaw = 0) position.  The profiles of all three gratings were identical within measurement uncertainty.  For the 100 $\mu$m slit configuration for the central CDB, where the line profile is the narrowest, a simple Gaussian model was fit to each separate data set and the width parameters were converted into FWHM.  The FWHMs in arcsec are $X4: 0.50\pm0.02''$, $X1: 0.45\pm0.06''$, and $X7: 0.47\pm0.04''$.  Compared to $0.47\pm 0.01''$ for the unobstructed beam we see no broadening from the gratings.

To compare scattering from the gratings we looked at the wings for the central 5 CDBs in the 100 $\mu$m slit configuration. For quantitative comparison we take the ratio of flux in the range 5-75 pixels from the peak (in the dispersion direction) to that within 75 pixels of the peak, similar to the analysis for the mirror line response in Fig.~\ref{unobstrWings}. The RMS variation between the three gratings in this flux fraction is 1.0\%, as compared with 0.5\% RMS of measurement errors. 

The zero orders have higher scattering than the unobstructed case.  However, the amount of increased scatter is only 1.5\% (X4, X7) and 3.5\% (X1), so the contribution to the HPW is still small.  For example, a 1" HPW optic would have a 1.07" zero order HPW if the grating contribution to scattering is 3\%.

\subsubsection{Zeroth order line response function (LRF) model}

\begin{figure}
 \begin{center}
   \includegraphics[width=3.4in]{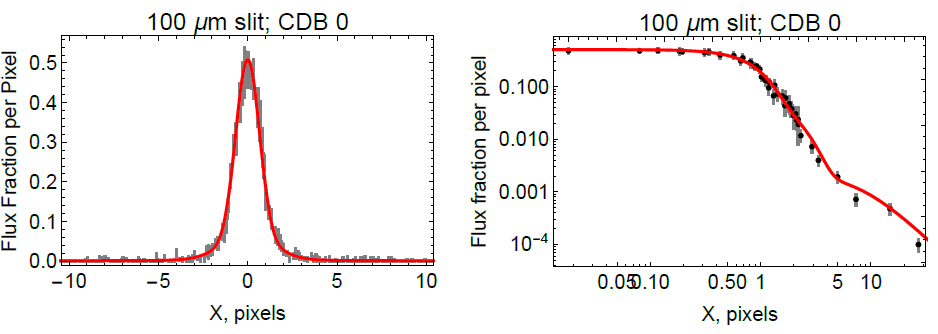}
 \end{center}
 \vspace{-0.3in}
\caption{\small Example best-fit LRF model (solid line) compared with data for the $100\,\mu{\rm m}$ slit, CDB 0 case. The right figure shows the absolute value of $X$ plotted against the flux fraction per pixel on log-log scales to emphasize the scattered component.  Sparse wing data have been binned to produce $\sim10\%$ error bars, and $X$ positions are the centroids of data within each bin. %
}
 \label{ZOfit}
\end{figure}

We developed an empirical model for the zeroth order line response function (LRF) for the source/optic/grating/detector system, consisting of the sum of contributions of two Gaussians of different widths to describe the central core and a Lorentzian function (Cauchy distribution) to account for scattering.  The model includes integration over the slit width and the pixel width. We simultaneously fit the data from 100 $\mu$m and 150 $\mu$m slit configurations in the five central CDBs. To simplify we combined data from the $\pm  1$ bands, as well as the $\pm 2$ bands. Fig.~\ref{ZOfit} compares the fit with data points for the center-most CDB. The log-log plot emphasizes the scattering wings. 
In Sec.~\ref{Sec4} we convolve the LRF with the expected source anode material line spectra to generate the predicted response for dispersed grating orders.

\subsection{Dispersed Line Profiles}

\begin{figure*}
 \begin{center}
   \includegraphics[width=7.0in]{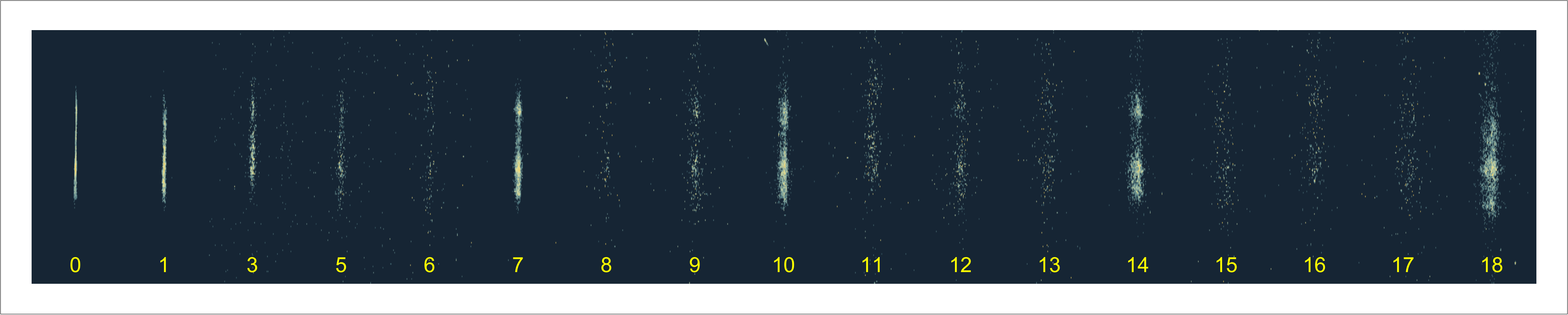}
 \end{center}
 \vspace{-0.15in}
\caption{\small Composite re-binned image of all orders collected using the Al anode. Orders 7, 10, 14, and 18 are long integrations. To make the images more visually comparable, pixel brightness is rescaled to the peaks of each re-binned image.  No images were taken at orders 2 and 4. First order corresponds to the resolving power of the Chandra HETG at this wavelength.}
 \label{AlOrderCollage}
\end{figure*}

\begin{figure}
 \begin{center}
   \includegraphics[width=3.5in]{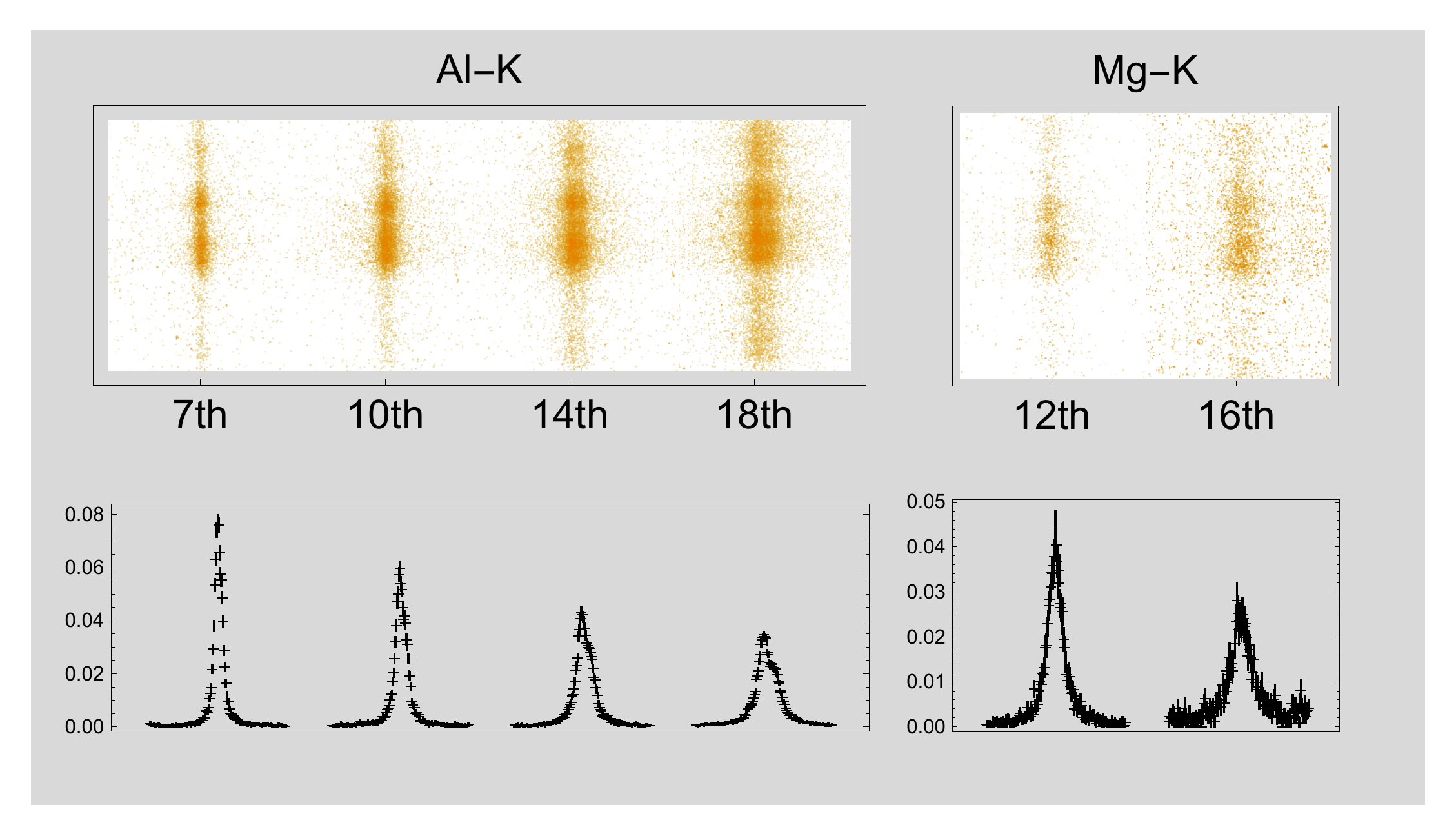}
 \end{center}
 \vspace{-0.15in}
\caption{\small Composite image of deep integration diffraction peaks collected for orders 7, 10, 14, and 18 using the Al anode, and 12 and 16 using the Mg anode. Pixel brightness has been rescaled, but images are at full resolution (no binning).  Line profiles are derived from simple column sums and normalized to unity.}
 \label{deepOrders}
\end{figure}

We collected CCD images at numerous dispersed orders of the K$_{\alpha}$ doublets using both the Al and Mg anodes.  In order to maximize the count rate for each measured order we want to align the grating for most efficient blazing for each order separately.  Blazing is strongest for orders under the blaze envelope, which is centered on the direction of specular reflection from the grating bar sidewalls, and has an angular width described well by the distance between minima of diffraction $w$ from a single slit of width $a$, the gap between bars, as $w \approx 2\lambda /a$.\cite{OE,SPIE2009,AO}
Therefore the gratings were rotated in yaw from normal incidence by half the angle of the dispersed order in an effort to maximize blazing.  For each order the camera was also translated along the optical axis relative to the 0$^{\rm th}$ order best focus to follow the expected best focus position (see Fig.~\ref{fig:layout}).
Fig.~\ref{AlOrderCollage} 
is a collage of all the orders for Al measured on one side of the 0$^{\rm th}$ order, scaled to a common maximum.  For Al-K x rays orders 7, 10, 14, and 18 were integrated for long times, while for Mg-K we integrated longer for orders 12 and 16 (not shown).  By far the longest integration was for 18$^{\rm th}$ order Al at 1726 minutes.  It is easily seen that higher orders have progressively broader peaks, making it easier to observe spectral features, and to deconvolve the source spectrum from the LRF.  If we simply sum along detector columns we obtain the line profiles shown in Fig.~\ref{deepOrders}.  Al orders 14 and 18 clearly show the well-known K$_{\alpha 1, \alpha 2}$ 2:1 intensity-ratio double peak shape.  The Mg anode produced much weaker flux than the Al one, and the data thus is noisier for the Mg peaks for similar integration times.

\section{Line Models and Grating Effective Resolving Power Analysis}
\label{Sec4}

To estimate grating effective resolving power, we modeled the fluorescent lines emitted by source anodes using literature values for measured line widths and wavelengths and applied the grating equation. When convolved with the zero order line profiles, the resulting diffracted profiles constitute a measurement prediction for an ideal grating.  Observed deviations from this model are interpreted as grating-induced. We initially model grating-induced broadening as a Gaussian grating period distribution.

\subsection{Line Models}

\begin{figure}
 \begin{center}
   \includegraphics[width=3.5in]{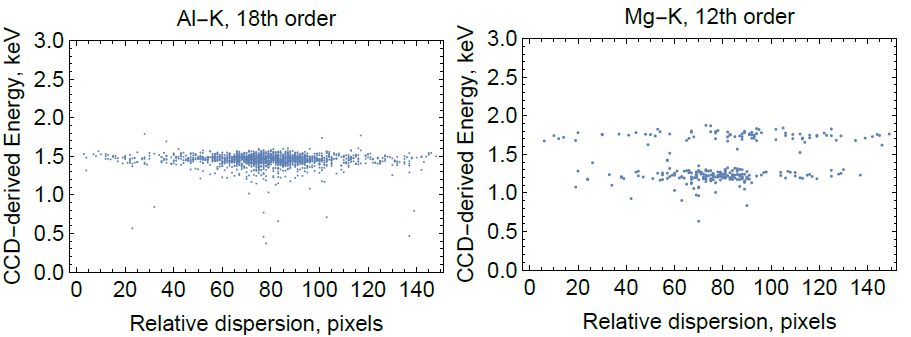}
 \end{center}
 \vspace{-0.15in}
\caption{\small   Order separation plots of dispersion vs. energy for nominal grade-zero \cite{cxc} 18$^{\rm th}$order Al-K and 12$^{\rm th}$order Mg-K events. The Mg-K data is contaminated with ~1.7-1.8 keV photons corresponding to 17$^{\rm th}$order W-$M_{\alpha_1}$ at 1.7754 keV or 0.6983 nm. The W-M line is much broader than the Mg-K lines (see Table \ref{lineTable}). Energy is determined from CCD collected charge using a single photon detection algorithm.}
 \label{ordSep}
\end{figure}

Long exposures were performed for diffracted orders 10, 14, and 18 for Al-K, and 12$^{\rm th}$ order for Mg-K.  The line purity from the Al anode did not require additional modeling beyond the expected K$_{\alpha 1}$ and K$_{\alpha 2}$ lines. This was not the case for the Mg anode.  Fig.~\ref{ordSep} presents an order separation analysis, where the energy resolution of the detector is used to distinguish between photons of different wavelengths and diffraction orders that have the same dispersion angle.  While the 18$^{\rm th}$ order Al data appears monochromatic, the 12$^{\rm th}$ order Mg data is contaminated with 17$^{\rm th}$ order W-M$_{\alpha 1}$ photons, probably due to W deposition on the Mg target from the heated cathode tungsten wire.  In addition, we need to include Mg-K line energies from both Mg and MgO to attain a satisfactory fit that was consistent with the Al-K results. The existence of both Mg and MgO was deduced from a fit which included the Mg K$_{\alpha 3}$ and K$_{\alpha 4}$ satellite lines, whose separation from each other and from the K$_{\alpha 1}$ and K$_{\alpha 2}$ lines differs significantly between the two cases \cite{fischer}.  Due to these complications and the lower Mg count rates, most of our analysis is focused on the Al data.

\begin{figure}
 \begin{center}
   \includegraphics[width=3.3in]{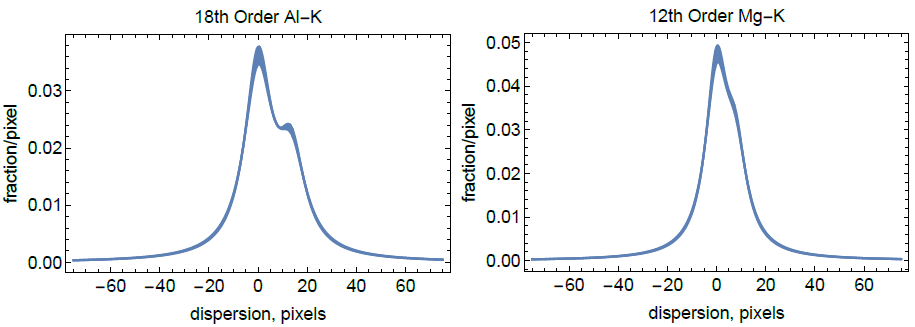}
 \end{center}
 \vspace{-0.2in}
\caption{\small   Predicted Al and Mg $K_{\alpha1}$ and $K_{\alpha2}$ lines without instrument effects. The plotted trace widths represent $\pm1\sigma$ uncertainties in the line separation and characteristic line widths, based on Table \ref{lineTable}.}
 \label{natLines}
\end{figure}

For reference we compiled Al-K, Mg-K, and W-M line energies and widths from various references in Table \ref{lineTable}. We included Al$_2$O$_3$ in the Table even though no evidence of additional Al lines was found.  In Fig.~\ref{natLines} we compare pure Al-K and Mg-K doublet line profiles based on values and uncertainties from Table \ref{lineTable}.   We are constrained in trying to deduce the CAT grating effective resolving power by the intrinsic line width uncertainty, regardless of the precision of the measurement.  For example, from Ref.~\cite{campbell} the intrinsic resolution of the Al-K lines is 3540 with line width uncertainties of 5\%.  A grating which causes the measured line width to exceed the natural line width by 5\% would have an effective resolving power of 16000 (using an empirical approximation to the Voigt function from Ref.~\cite{olivero}).  Attempting to attribute a deviation from the known value to uncertainty in the known value or to grating performance, we cannot rely on broadening alone, but need to discriminate between profile shapes.

\begin{table*}[h]
   \centering
{\footnotesize 
\begin{tabular}{ccccccccc} \hline 
\textbf {Material} & \textbf{Lines} & \multicolumn{2} {c} {\textbf{$\alpha_1$}} & \multicolumn{2} {c} {\textbf{$\alpha_2$}} & {\textbf{$\alpha_3$}} & {\textbf{$\alpha_4$}} & ref.\\ 
 & & \textbf{$\lambda$} & \textbf{$\Gamma$} & \textbf{$\lambda$} & \textbf{$\Gamma$} & \textbf{$\lambda$} & \textbf{$\lambda$}  \\ \hline \hline
 &&&&&&&&{\vspace{-0.07in}}\\
 \text{Al} & \text{Al-K} & $0.8339527$ &  $0.4\pm0.1$ & $0.8341843$ &  $0.4\pm0.1$ &  &   & \cite{schweppe}(1994)\\
  & & $\pm0.0000056$ &  & $\pm0.0000056$ &  &  &   & \\
 \text{Al} & \text{Al-K} & $0.833934$ &   & $0.834173$ &   &  &   & \cite{bearden}(1967)\\
 & & $\pm0.000009$ &   & $\pm0.000009$ &   &  &    & \\
 \text{Al} & \text{Al-K} &  &   &  &   & $0.82854$ &   $0.82744$  & \cite{krause2}(1975)\\
 &  &  &   &  &   & $\pm0.00006$ &   $\pm0.00006$ &   \\
 \text{Al} & \text{Al-K} & 0.83393 &   &  &   & $0.82854$ &   $0.82744$   & \cite{baun}(1964)\\
 &   & $\pm0.00006$ &   &  &   & $\pm0.00006$  & $\pm0.00006$   & \\
 \text{${\rm Al}_2{\rm O}_3$} & \text{Al-K} & 0.83380 &   &  &   & $0.82820$ &   $0.82718$ &   \cite{baun}(1964)\\
 &  &  $\pm0.00006$ &   &  &   & $\pm0.00006$ &   $\pm0.00006$ &   \\
\text{Al} & \text{Al-K} &   & $0.42\pm0.02$ & & $0.42\pm0.02$ &  &  &  \cite{campbell}(2001)\\
 \text{Al} & \text{Al-K} &   & $0.43\pm0.04$ & & $0.43\pm0.04$ &  &  &   \cite{krause}(1979)\\
 \text{Mg} & \text{Mg-K} & $0.9889573$ & $0.4\pm0.1$ & $0.9891553$ & $0.4\pm0.1$ &  &  &   \cite{schweppe}(1994)\\
  & & $\pm0.0000087$ &  & $\pm0.0000103$ &  &  &  &  \\
 \text{Mg} & \text{Mg-K} & $0.98900$ &  & $0.98900$ &  &  &  &  \cite{bearden}(1967)\\
  & & $\pm0.00002$ &  & $\pm0.00002$ &  &  &  & \\
 \text{Mg} & \text{Mg-K} &  &  &  &  & $0.98230$ &  $0.98105$ &   \cite{krause2}(1975)\\
  & &  &  &  &  & $\pm0.00008$   & $\pm0.00008$ &   \\
 \text{Mg} & \text{Mg-K} & 0.98912  &  &  &  & $0.98248$ &   $0.98123$ &   \cite{fischer}(1965)\\
  & & $\pm0.00008$ &  &   &  & $\pm0.00008$ &   $\pm0.00008$ &   \\
 \text{MgO} & \text{Mg-K} & 0.98889  &  &  &  & $0.98210$  & $0.98080$ &   \cite{fischer}(1965)\\
  & & $\pm0.00008$ &  &   &  & $\pm0.00008$ &   $\pm0.00008$ &   \\
  \text{Mg} & \text{Mg-K} &   & $0.36\pm0.02$ &  & $0.36\pm0.02$ & &    & \cite{campbell}(2001)\\
 \text{Mg} & \text{Mg-K} &   & $0.36\pm0.04$ &  & $0.36\pm0.04$ & &    & \cite{krause}(1979)\\
 \text{W} & \text{W-M} & $0.6983$ &  & $0.6992$ &  &  &    & \cite{bearden}(1967)\\
 & & $\pm0.0001$ &  & $\pm0.0002$ &  &  &   & \\
 \text{W} & \text{W-M} &   & $1.76\pm0.18$ &  & $1.80\pm0.18$ & &  & \cite{campbell}(2001)\vspace{-0.1in}\\
\end{tabular}
}
  \caption{\small \textbf{X-ray lines.}
Line wavelengths, $\lambda$, (nm) and widths, $\Gamma$, (eV) from various references. }
   \label{lineTable}
\end{table*}

The K$_{\alpha}$ doublets are modeled in the following manner: We assume that the naturally-occurring energy distribution of photons around the line energies follows a Lorentzian function or Cauchy distribution described by:
\begin{equation}
L(x,a,b)=\frac{1}{\pi b \left(\frac{(x-a)^2}{b^2}+1\right)},  \label{eq:lor1}
\end{equation}
with ${\rm FWHM}=f_L=2b$, and peak at $x=a$.
For comparison with data we integrate $L$ across a pixel width, which is the scale at which we bin the integrated charge. Operationally, the function used is:
\begin{equation}
L_1(x,a,b,d)=\frac{\tan ^{-1}\left(x-a+\frac{d}{2} \right)-\tan ^{-1}\left(x-a-\frac{d}{2} \right)}{\pi},  \label{eq:lor2}
\end{equation}
where $d$ is the width of the integration bin in the same units as $a$ and $b$. We maintain pixels as the dispersion distance units throughout the analysis, converting only for certain figures. The line models we define are local in wavelength to the Al and Mg K$_{\alpha1}$ so it is only necessary to specify wavelength differences relative to these. We have extracted nominal values for all the necessary physical constants from Table \ref{lineTable} and listed them in Table \ref{lineTable2}. The modeled width and separation parameters are normalized relative to these nominal values.  Setting all of these model parameters to 1.0 gives the nominal model.

\begin{table*}[h]
   \centering
{\footnotesize \begin{tabular}{cclc} \hline 
\textbf{Model Constants} & \textbf{Nominal Values} & \textbf{Description} & \textbf{Ref.}  \\ \hline \hline
 &&&{\vspace{-0.07in}}\\
$\Gamma^{\tiny {\it A}}_{\alpha1}=\Gamma^{\tiny {\it A}}_{\alpha2}$ &\text{236 fm, 0.42 eV}&\text{Al $K_{\alpha1}$,$K_{\alpha2}$ Lorentzian FWHM}& \cite{campbell}\\
$\Gamma^{\tiny {\it B}}_{\alpha1}=\Gamma^{\tiny {\it B}}_{\alpha2}=\Gamma^{\tiny {\it C}}_{\alpha1}=\Gamma^{\tiny {\it C}}_{\alpha2}$ &\text{284 fm, 0.36 eV}&\text{Mg $K_{\alpha1}$,$K_{\alpha2}$ Lorentzian FWHM for both Mg and MgO}&\cite{campbell}\\
$\Gamma^{\tiny {\it D}}_{\alpha1}$ &\text{692 fm, 1.76 eV}&\text{W $M_{\alpha1}$ Lorentzian FWHM}&\cite{campbell}\\
$\lambda^{\tiny {\it A}}_{\alpha2}-\lambda^{\tiny {\it A}}_{\alpha1}$ &\text{232 fm (0.413 eV)}&\text{Al $K_{\alpha1}$,$K_{\alpha2}$ separation}&\cite{schweppe}\\
$\lambda^{\tiny {\it B}}_{\alpha2}-\lambda^{\tiny {\it B}}_{\alpha1}=\lambda^{\tiny {\it C}}_{\alpha2}-\lambda^{\tiny {\it C}}_{\alpha1}$ &\text{198 fm (0.251 eV)}& \text{Mg $K_{\alpha1}$,$K_{\alpha2}$ separation for both Mg and MgO}&\cite{schweppe}\\
$\lambda^{\tiny {\it C}}_{\alpha1}-\lambda^{\tiny {\it B}}_{\alpha1}$ &\text{-230 fm (0.292 eV)}& \text{Mg $K_{\alpha1}$ for pure Mg separation from Mg $K_{\alpha1}$ for MgO}&\cite{fischer}\\
$\frac{17}{12} \lambda^{\tiny {\it D}}_{\alpha1} - \lambda^{\tiny {\it B}}_{\alpha1}$ &\text{300 fm}& \text{12$^{\rm th}$Mg $K_{\alpha1}$ for pure Mg separation from 17$^{\rm th}$W $M_{\alpha1}$}&\cite{schweppe},\cite{bearden}\\
\end{tabular}}
  \caption{\footnotesize \textbf{Nominal as-modeled x-ray line widths and separations.}
Superscripts are defined as A-(Al lines), B-(Mg lines from pure Mg), C-(Mg lines from MgO), D-(W lines).  We have assumed the equivalencies listed in the first column. The wavelength unit, {\it fm}, is femto-meters ($10^{-15}$m). For comparison, 1 pixel at orders $\{7,10,12,14,18\}$ extends over $\{44.05, 30.83, 25.70, 22.02, 17.13\}$ fm in wavelength space.}
   \label{lineTable2} 
\end{table*}

Our model for the $i$$^{\rm th}$order Al $K_{\alpha1}$,$K_{\alpha2}$ lines is:
\begin{eqnarray}
{\bf Al}_0( x, i , x_0, L_{\rm Al}, S_{\rm Al}, d ) & = & (2/3) L_1(x, x_0,\frac{f_{\alpha1}}{2},d)+\nonumber \\
& & + (1/3)L_1(x, x_0+ \Delta_{\alpha1,2} ,\frac{ f_{\alpha2}}{2},d) \nonumber \\
f_{\alpha1} & = & i \kappa L_{\rm Al} \Gamma^{\tiny {\it A}}_{\alpha1} \nonumber \\
f_{\alpha2} & = & i \kappa L_{\rm Al}\Gamma^{\tiny {\it A}}_{\alpha2} \nonumber \\
\Delta_{\alpha1,2} & = & i \kappa S_{\rm Al}(\lambda^{\tiny {\it A}}_{\alpha2}-\lambda^{\tiny {\it A}}_{\alpha1}), \label{eq:al0}
\end{eqnarray}
\noindent where $x$ is the dispersion axis coordinate, $x_0$ is the center position of Al $K_{\alpha1}$, $L_{\rm Al}$ is the fraction of the nominal Lorentzian FWHM, $S_{\rm Al}$ is the fraction of the nominal line separation, and $d$ is the bin width in units defined by $\kappa$.  $f_{\alpha1}$ and $f_{\alpha2}$ are as-modeled Lorentzian FWHMs for the two lines, which are assumed equal. $\kappa$ is a scale factor to convert wavelength to detector distance. $\Gamma^{\tiny {\it A}}_{\alpha2}$ and $(\lambda^{\tiny {\it A}}_{\alpha2}-\lambda^{\tiny {\it A}}_{\alpha1})$ are defined in Table \ref{lineTable2}. The normalization yields 1.0 for the sum of points calculated at interval $d$. We have assumed the flux ratio for $K_{\alpha1}$/ $K_{\alpha2}$ is 2.

Models for other lines (Mg, MgO, W) follow equivalent expressions.

\subsection{Measurement Model}

The measurement model relates the line models to the measured integrated charge profile detected by the camera. The ingredients are the modeled line shapes, the grating dispersion relation, the zeroth order LRF, and the modeled dispersed grating response, assumed to be Gaussian.

We define a measurement model ${\bf Al}_\rho$, where $\rho$ identifies the cross-dispersive region of interest in terms of the number of CDBs as defined in Sec.~\ref{meas}\ref{sourcemirror}. We combine data from various CDBs cumulatively in terms of radius from the center. Three values of $\rho$ are considered initially:
\begin{eqnarray}
 \rho=1 & &\text{CDB:$\{0\}$} \nonumber \\
 \rho=3 & &\text{CDB:$\{0,\pm1\}$} \nonumber \\
 \rho=5 & &\text{CDB:$\{0,\pm1,,\pm2\}$.} \nonumber
\end{eqnarray}
To go from function ${\bf Al}_0$ to ${\bf Al}_\rho$, we scale ${\bf Al}_0$ to the proper dispersion units, then convolve it with the appropriate zero order LRF for the given $\rho$, and finally convolve it with a normalized Gaussian (FWHM $f_G$) representing the dispersed grating response. 

We derive the dispersion relation in two ways, first from the known grating period and measured grating-to-detector distance, and second from the detector stage translation and CCD image positions (horizontal and vertical, since the dispersion axis was slightly off from horizontal).
For the first, with 200 nm grating period and a distance of 8750 mm we get 43.750 mm/nm.  For 18$^{\rm th}$ order Al-K we measured a separation from zeroth order of 657.15 mm, which gives 43.777 mm/nm in first order.
The uncertainty in the first value is dominated by the distance error of $\pm5$ mm giving an uncertainty of $\pm0.025$ mm/nm. The separation uncertainty is $\pm0.1$ mm based on the known positioning repeatability of the CCD stages, which gives $\pm0.007$ mm/nm uncertainty for the second value.  We therefore use 43.777 mm/nm or $\kappa=3242.74$ pixels/nm, and conclude that the $0.02\%$ error is negligible, especially compared with at least $\sim5\%$ uncertainty in the natural line widths.  We also use the notation for dispersion distance $D_{mi}= i \kappa \lambda_m$, where $i$ is the grating order, $m$ is the anode material (Al or Mg), and $\lambda_m$ is the $K_{\alpha 1}$ wavelength for the material.

\begin{figure}
   \includegraphics[width=3.0in]{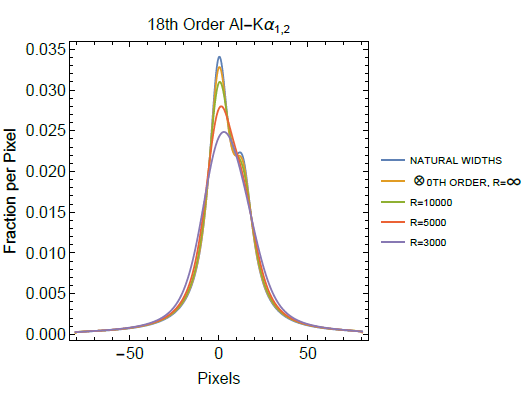}
   \includegraphics[width=3.0in]{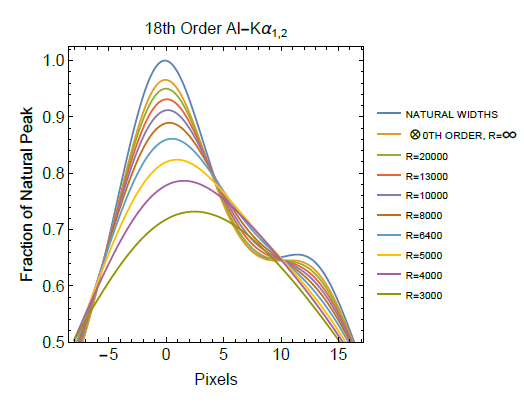}
 \vspace{-0.2in}
\caption{\small  Illustration of the measurement model for 18$^{\rm th}$ order Al. Top panel shows the full profile at coarsely sampled effective resolving powers. Bottom panel shows the core region at finely sampled effective resolving powers. }
 \label{AlModel}
\end{figure}

Fig.~\ref{AlModel} indicates the effect of the zero order LRF and various levels of $f_G$ on the line profile for 18$^{\rm th}$ order Al-K.

\subsection{Grating Effective Resolving Power Analysis}
We performed independent fits to individual diffraction orders to estimate $f_G$ and resolving power, as well as fits to ensembles of orders from a single source, and ensembles of orders from both Al and Mg combined.  Fit analysis for 18$^{\rm th}$ order Al is described in detail.  Analysis of other orders and ensembles is done in similar fashion.
\subsubsection{18$^{\rm th}$ Order Al-K.}
The 18$^{\rm th}$ order Al-K data was fit in two stages: first with only $f_G$ variable, and then with all parameters variable.
In the first stage, we separately fit the full profile  and the core region interior to the FWHM, to explore possible systematic biases.
For all of these cases, the model is:
\begin{equation}
Q_\rho(x; \Phi)  =  Q_0+ n  {\bf Al}_\rho \left(18,L_{\rm Al}, S_{\rm Al}, f_G  \right) \left( x-x_{0} \right)  \\ \label{eq:Almodel}
\end{equation}
\noindent with parameters listed in Table \ref{AlparamTable}. $Q_0$ represents any continuum from other orders. ${\bf Al}_\rho$ is the measurement model function for $\rho$ CDBs, defined in the previous paragraphs.

\begin{table}[h]
   \centering
{\scriptsize \begin{tabular}{cccl}\hline 
\text{Parameter} & \text{Tab. \ref{AlKfixedfit}(a-f)} & \text{Tab. \ref{AlKfreeFit}(a,b)}& \text{Description} \\ \hline \hline
$Q_0$ & \text{var} & \text{var} &  \text{signal floor} \\
$n$  & \text{var} & \text{var} &  \text{18$^{\rm th}$order Al normalization} \\
$f_G$  & \text{var} & \text{var} & \text{Gaussian FWHM for 18$^{\rm th}$order} \\
$L_{\rm Al}$  & 1.0 & \text{var} &  \text{fraction of nominal Al-$K\alpha$ natural line width} \\
$S_{\rm Al}$  & 1.0 & \text{var} & \text{fraction of nominal Al-$K\alpha_{1,2}$ line separation} \\
$x_0$  & \text{var} & \text{var} &  \text{18$^{\rm th}$order Al-$K\alpha_1$ peak $x$ position} \\
\end{tabular}}
  \caption{\small \textbf{Ensemble fit parameter summary.} The list forms the set of parameters for Eq. \ref{eq:Almodel}.  Fixed parameter values are listed in column two for the fit cases defined in the table referenced in the top row.}
   \label{AlparamTable}
\end{table}

The spectrum was first fit with all wavelengths and line widths fixed at their nominal values from Table \ref{lineTable2}, but with variable Gaussian FWHM. Best fit models for the $\rho=1$, 3 and 5 cases were compared, using wide (143 pixels) and narrow (27 pixels) regions of interest (ROI) in the dispersion direction. We performed a $\Delta \chi ^2$ analysis with 1 d.o.f., varying $f_G$.  Key parameters of these fits are listed in Table \ref{AlKfixedfit}.  The resulting $\chi^2$ values indicate good fits. The last column in Table \ref{AlKfixedfit}, $P(\chi^2)$, tabulates the values of the cumulative $\chi^2$ probability for the given degrees of freedom at the measured $\chi^2$ value from the weighted fit residuals.

\begin{table}[h]
   \centering
{\tiny \begin{tabular}{ccccccccc}\hline 
\textbf{Fit ID} &\textbf{CDB} & \textbf{ROI} & \textbf{$f_G$} & \textbf{$R_g$ (90$\%$ l.l.)} & \textbf{$R_g$ (90$\%$ l.l.)} & $\chi^2$ & \text{DOF} & $\left.\text{P(}\chi^2\right)$ \\
 \text{} & \text{} & \text{pixels} & \text{pixels} & \text{from }$\Delta \chi ^2$ & \text{from fit} & \text{} & \text{} & \text{} \\\\ \hline \hline
 \text{(a)} &  0 & 143 & \text{0+2.8} & 15600 & - & 125.0 & 135 & 0.28 \\
 \text{(b)} & \text{0,$\pm $1} & 143 & \text{0+1.8} & 24500 & - & 133.6 & 135 & 0.48 \\
 \text{(c)} &  \text{0,$\pm $1,$\pm $2} & 143 & \text{0+2.0} & 21500 & - & 126.0 & 135 & 0.30 \\
 \text{(d)} &  0 & 27 & \text{3.3$\pm $1.8} & 9100 & 8700 & 20.55 & 19 & 0.64 \\
 \text{(e)} &  \text{0,$\pm $1} & 27 & \text{2.1$\pm $1.5} & 12600 & 12200 & 16.76 & 19 & 0.39 \\
 \text{(f)} &  \text{0,$\pm $1,$\pm $2} & 27 & \text{2.6$\pm $1.2} & 12100 & 12200 & 17.62 & 19 & 0.45 \\
 \end{tabular}}
  \caption{\small \textbf{18$^{\rm th}$ order Al summary of fits with fixed line wavelengths and widths.}
Rows (a)-(c) had a wide ROI, (d)-(f) had a narrow ROI. Effective resolving power values, $R_g$, use dispersion distances $D_{\rm Al18}$. (90$\%$ l.l.) means 90$\%$ probability that $R_g$ is greater than the given value.}
   \label{AlKfixedfit}
\end{table}

Using the wide ROI (rows (a)-(c) in Table \ref{AlKfixedfit}) we find that the lowest $\chi^2$ is at $f_G=0$, or $R_g=\infty$ with $1\sigma$ errors in the 2-3 pixel range. Using the narrow ROI (rows (d)-(f)) the $\chi^2$ is a minimum in the 2-3 pixel range, but errors are large, 50-75$\%$. The two cases are consistent with one another, with the narrow ROI being less sensitive, since it contains less information. In both ROI cases the $\rho=1$ fits, (a) and (d), had larger error due to  lower count rates.  For this reason we focus on analysis of $\rho=3$ and 5 data in the rest of this work.  But even from the data with the most counts (cases (b) and (c)) we cannot simply conclude the grating has better than 20000 effective resolving power,
since $L_{\rm Al}$ is unknown at the $5\%$ level and $S_{\rm Al}$ is unknown at the $3\%$ level, and these parameters are somewhat correlated with $f_G$.  

\begin{figure}
 \begin{center}
   \includegraphics[width=3.5in]{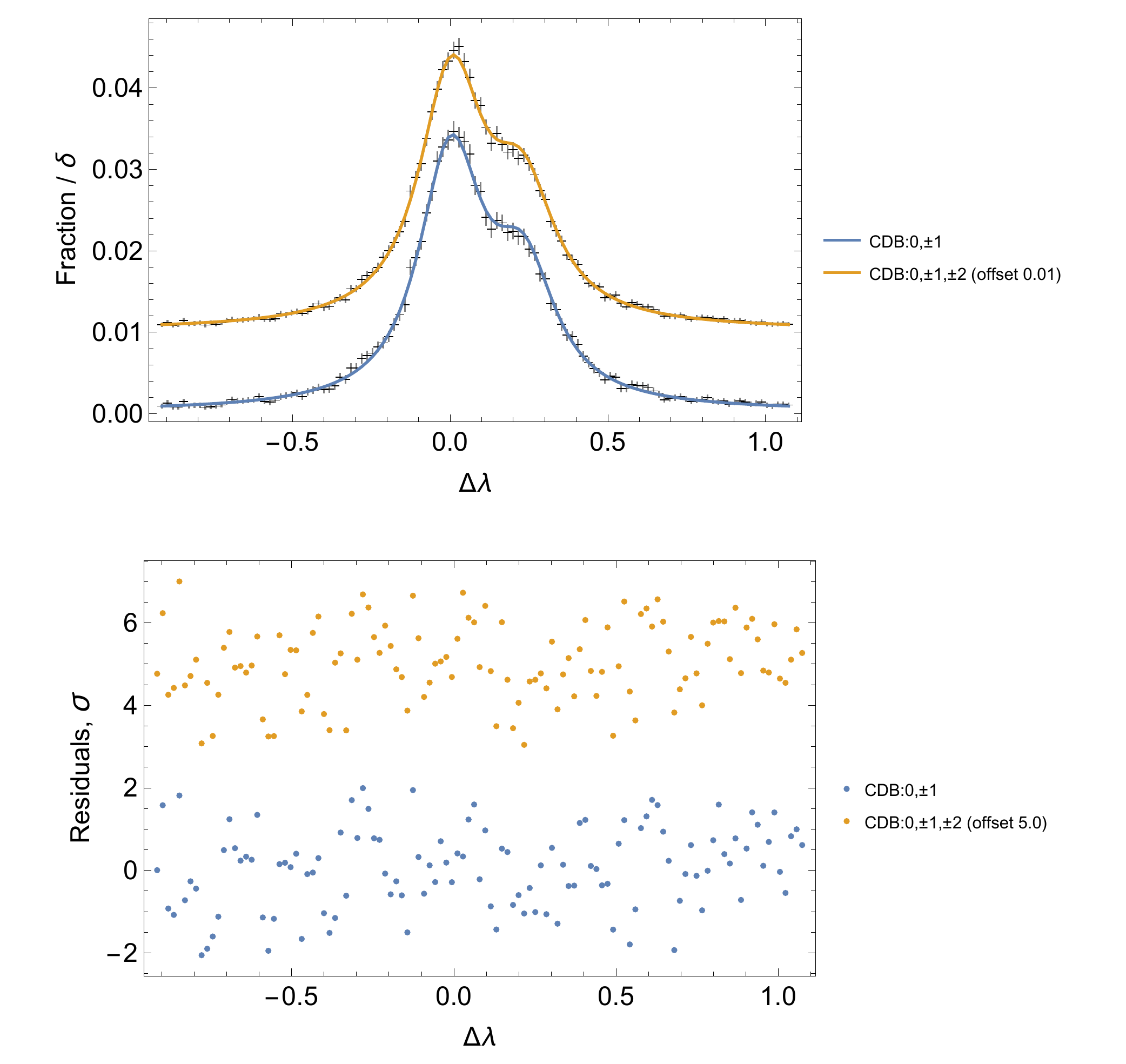}
 \end{center}
 \vspace{-0.2in}
\caption{\small  18$^{\rm th}$ order Al best fit models, for 3 and 5 CDB cases compared with data. Key fit results are listed in Table \ref{AlKfreeFit} under Fit IDs (a) and (b). $\Delta \lambda = 0$ corresponds to $\lambda = 833.95$ pm.}
 \label{AlFitFreeWd}
\end{figure}

\begin{figure}
 \begin{center}
   \includegraphics[width=3.0in]{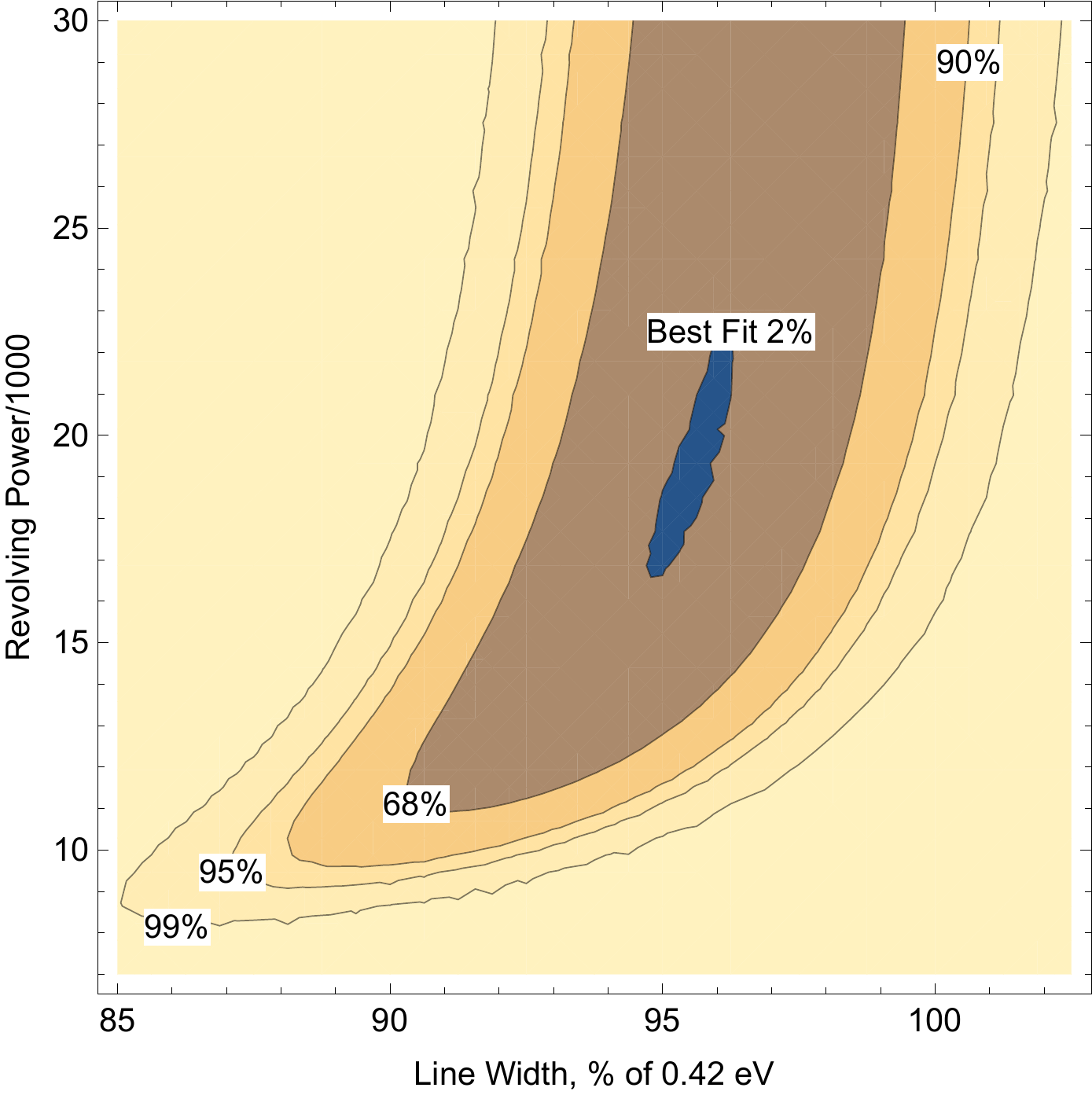}
   \includegraphics[width=3.0in]{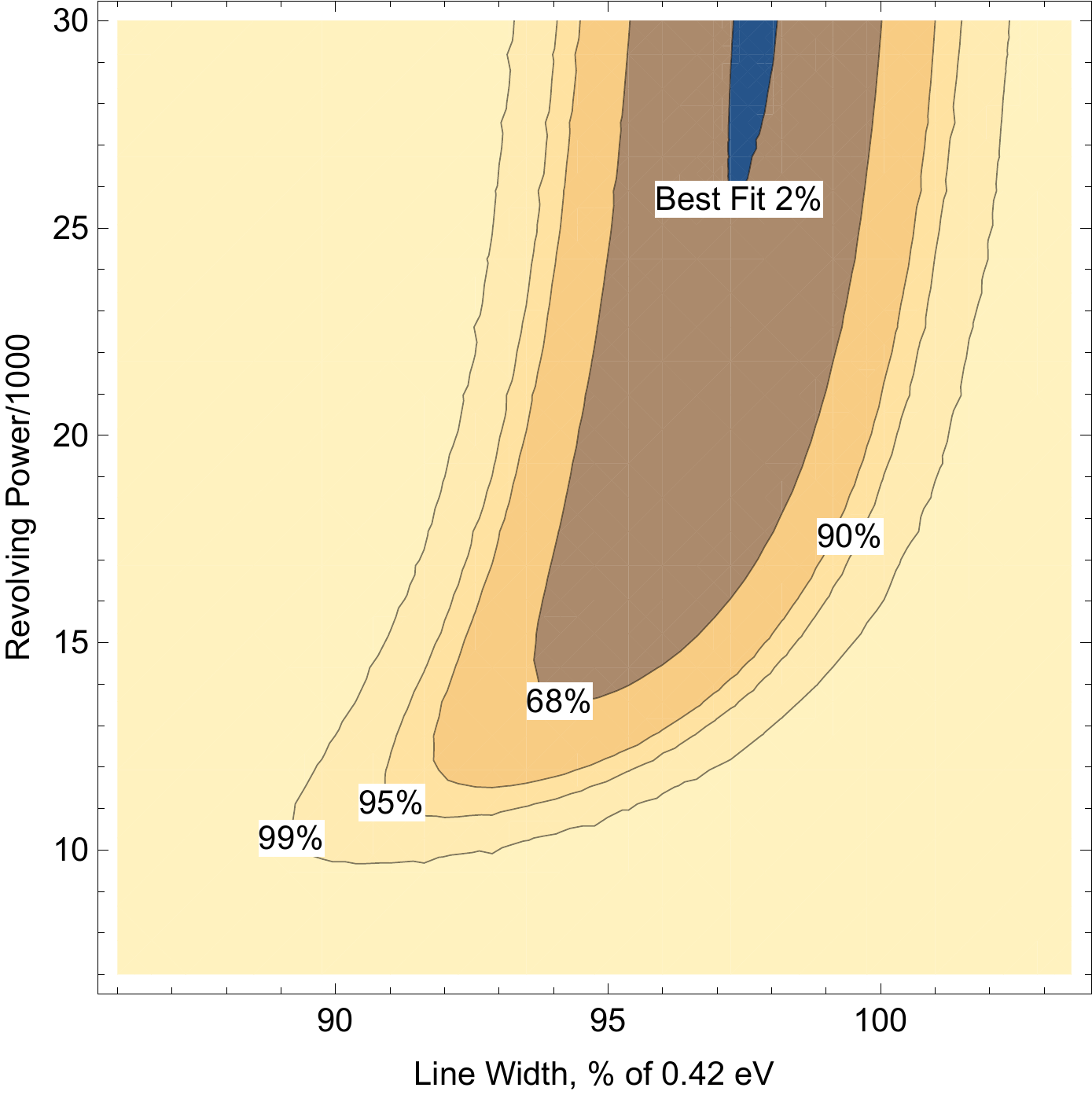}
 \end{center}
 \vspace{-0.2in}
\caption{\small  Results of $\Delta \chi^2$ analysis for 18$^{\rm th}$order Al. [Top, Bottom] are probability contours for $\rho=[ 3, 5 ]$ CDB cases.  The blue area represents the 2\% probability around the best fit values.  The value of 0.42 eV for the line width is taken from Table~\ref{lineTable}, Ref.~\cite{campbell}.}
 \label{AlFitcont1}
\end{figure}

Next, $L_{\rm Al}$ and $S_{\rm Al}$ are allowed to vary. Fig. \ref{AlFitFreeWd} compares best fit models for both $\rho=3$ and 5 cases with data.  
We calculated $\Delta \chi^2$ over a 2-dimensional grid, varying the Gaussian FWHM, $f_G$, and the Al line width parameter $L_{\rm Al}$ over specific grid values, while fitting the other parameters. Fig.~\ref{AlFitcont1} shows the $\chi^2$ probability contours for two degrees of freedom.  The $\rho=5$ case $95\%$ probability contours indicate $\sim15\%$ higher resolving power lower limits, $\sim11000$, than $\rho=3$, with $R_g=D_{\rm Al18}/f_G$. The contours also indicate the range of acceptable values for $L_{\rm Al}$ with best case for $\rho=5$ with a $1\sigma$ uncertainty of $\ ^{+2\%}_{-4\%}$. The contours are consistent with the nominal 0.42 eV line width, with best fit at $98\%$ of this value.  

The first two columns in Table \ref{AlKfreeFit} show the key information from these two fits.  The $\chi^2$ values suggest both are good fits.  The $R_g$ values from the fit errors and the $\Delta \chi^2$ differ by $5-20\%$. This is mainly because the parameter fit errors are determined from the diagonal elements of the covariance matrix of a linearized version of the model, whereas the $\Delta \chi^2$-derived errors account for asymmetries in the non-linear model. The fit parameter errors for $L_{\rm Al}$ were consistent with those obtained from the $\Delta \chi^2$ contours after accounting for similar asymmetry. $S_{\rm Al}$ errors were $<1.5\%$.

The column (b) result indicates that the grating effective resolving power lower limit is nearly 11000 at $95\%$ confidence limit (c.l.). The $90\%$ c.l. value for $\rho=5$ changed from 12100 to 11500, a $5\%$ decrease, as a result of allowing the  $L_{\rm Al}$ and $S_{\rm Al}$ to vary. This indicates that the Al data provides a robust measurement of the fit parameters.

\subsubsection{Other Al-K Orders.}
Profiles derived from deep integrations at Al 14$^{\rm th}$, 10$^{\rm th}$, and 7$^{\rm th}$ orders were also fit with all parameters variable using the same model as 18$^{\rm th}$ order with the obvious adjustments. Key fit parameters are summarized for $\rho = 3$ under Fit IDs (c), (e) and (g) in Table \ref{AlKfreeFit}. 14$^{\rm th}$ and 10$^{\rm th}$ order were also fit for $\rho=5$ CDB cases, and fit parameters are summarized under Fit IDs: (d) and (f) in Table \ref{AlKfreeFit}. 
Results for these cases and the two 18$^{\rm th}$ order fits are in good agreement.  

The Gaussian FWHM results are all consistent with the same value of approximately two pixels. Errors in $f_G$ are generally large with the smallest $\sim30\%$ for 7$^{\rm th}$order. Possible implications of these results are further discussed in Sec.~\ref{discuss}.  Of course, the cases are not all independent since the $\rho=5$ cases include the $\rho=3$ data. The best fit values for $L_{\rm Al}$ are also consistent with one another, as well as the nominal value.   The best fit values for $S_{\rm Al}$ are consistent with one another, but are generally less than one and in the case of 10$^{\rm th}$ order, are significantly less than one.  The resolving power best fit values and lower limits decline with order because of the decreasing dispersion distance, since the best fit $f_G$ and errors are relatively constant. Only the 7$^{\rm th}$ order case excludes $R_g=\infty$ at a significant level. The conclusion is as expected, that we are most sensitive to $R_g$ at the larger orders, and the 18$^{\rm th}$ order limits are the strongest constraint on the broadening resulting from grating period variation, $\Delta p/p$.  14$^{\rm th}$ and 10$^{\rm th}$ order best fit $f_G$ values still suggest very high resolution, so we next describe a simultaneous fit of combined 18$^{\rm th}$, 14$^{\rm th}$ and 10$^{\rm th}$ order data to check that the lower order results are consistent with the 18$^{\rm th}$ order data, and to try to achieve even better constraints on the fit parameters.

\subsubsection{Ensemble Fitting of Al-K Orders.} 
\begin{figure}
 \begin{center}
 \includegraphics[width=3.5in]{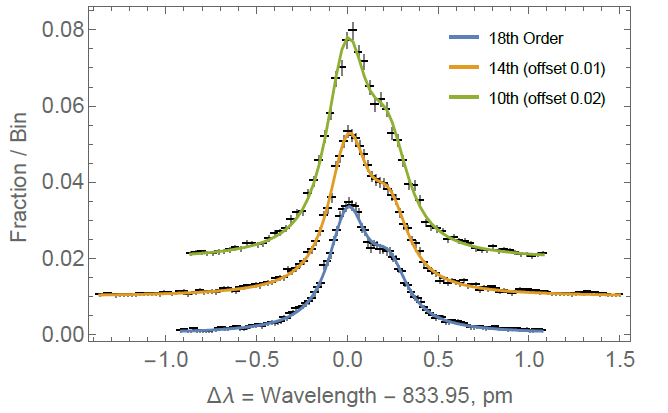}
 \end{center}
 \vspace{-0.2in}
\caption{\small  Al-(18$^{\rm th}$, 14$^{\rm th}$, 10$^{\rm th}$) combined best fit model for $\rho = 5$  case compared with data. Key fit results are listed in Table \ref{AlKfreeFit} under Fit ID (i).}
 \label{AlFitComb2}
\end{figure}

For the fit of the ensemble 18$^{\rm th}$, 14$^{\rm th}$ and 10$^{\rm th}$ order Al data each order was normalized independently.  We used a single Gaussian FWHM parameter, but scaled by dispersion distance relative to 18$^{\rm th}$ order for each order. Fig.~\ref{AlFitComb2} compares best fit models for the $\rho=5$ case with data.

\begin{figure}
 \begin{center}
 \includegraphics[width=3.0in]{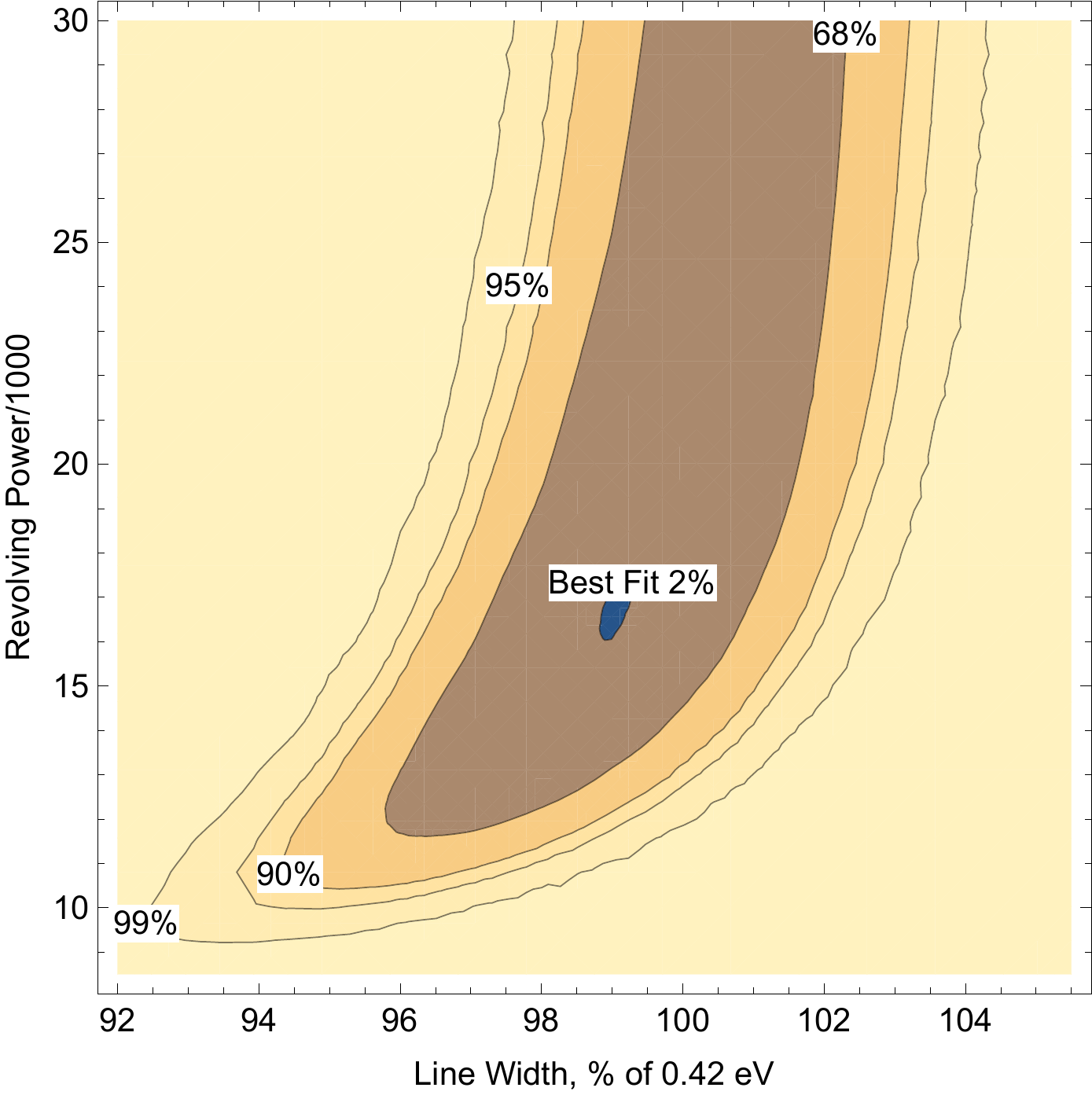}
 \includegraphics[width=3.00in]{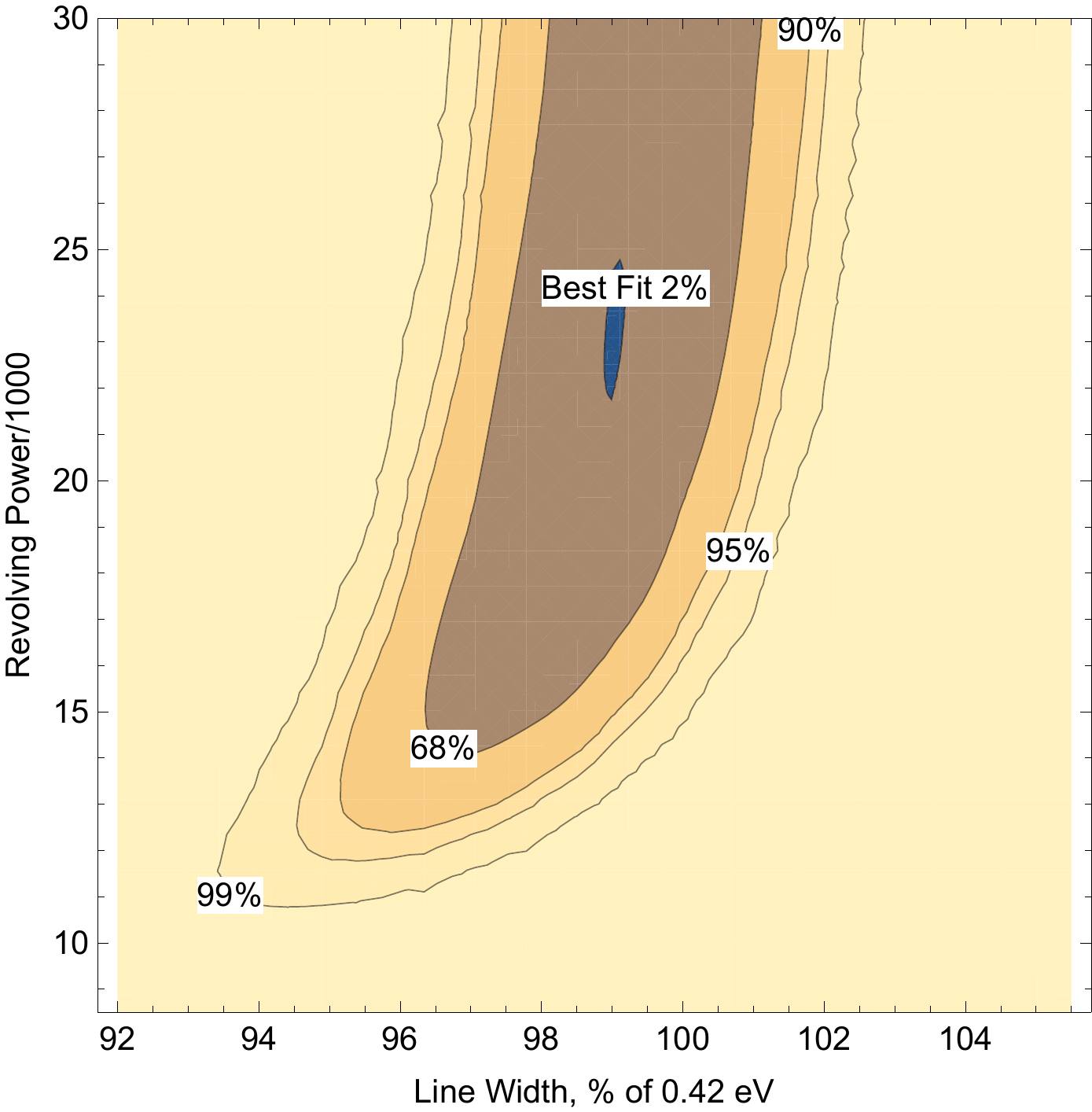}
 \end{center}
 \vspace{-0.2in}
\caption{\small  Results of $\Delta \chi^2$ analysis for combined Al-(18$^{\rm th}$, 14$^{\rm th}$, 10$^{\rm th}$). [Top, Bottom] are probability contours for $\rho=$[3, 5] CDB cases.}
 \label{AlFitCombCont}
\end{figure}

We again calculated $\Delta \chi^2$ over a 2-dimensional grid, varying the Gaussian FWHM, $f_G$, and the Al line width parameter $L_{\rm Al}$ over specific grid values, while fitting the other parameters. Fig.~\ref{AlFitCombCont} shows the $\chi^2$ probability contours for two degrees of freedom.  The 5 CDB case $95\%$ c.l.~contours indicate $\sim20\%$ higher resolving power lower limits, $\sim12000$, than $\rho=3$, and $\sim10\%$ higher than 18$^{\rm th}$ order alone. Because of our scaling of $f_G$ in the model, $R_g=D_{\rm Al18}/f_G$ for these fits .

The contours indicate a narrower range of acceptable values for $L_{\rm Al}$ than attained with only 18$^{\rm th}$ order.  The best case was the $\rho = 5$ case with $2\%$ $1\sigma$ uncertainty. The contours agree well with the nominal 0.42 eV line width, with best fit at $99\%$ of this value.  This appears to be an improvement in precision over the best previously quoted uncertainty in the line width from Ref.~\cite{campbell} (see Sec. \ref{discuss}).

The last two columns in Table \ref{AlKfreeFit} show the key information from these two fits.  The $\chi^2$ values suggest both are acceptable fits, though column (h) is the worst of all our fits with 0.93 cumulative $\chi^2$ probability.  The $R_g$ values from the fit errors and the $\Delta \chi^2$ are in reasonable agreement, differing by $<10\%$ due to asymmetries in the $\Delta \chi^2$ contours. The fit parameter errors for $L_{\rm Al}$ are consistent with those obtained from the $\Delta \chi^2$ contours, and $S_{\rm Al}$ errors are slightly smaller than those attained from 18$^{\rm th}$ order alone, $\sim1\%$.

The column (i) result indicates that the grating effective resolving power lower limit is nearly 12000 at $95\%$ c.l. This value is higher than that obtained with 18$^{\rm th}$ order alone, indicating that combining the three orders provides increased measurement sensitivity. 

\begin{table*}[h]
   \centering
{\scriptsize \begin{tabular}{lccccccccc}\hline 
\text{Fit ID} & \text{(a)} & \text{(b)} & \text{(c)} & \text{(d)} & \text{(e)} & \text{(f)} & \text{(g)} & \text{(h)} & \text{(i)}\\
\text{Order} &18 & 18 & 14 & 14 & 10 & 10 & 7 & 18-14-10 & 18-14-10 \\
\text{CDB} & \text{0,$\pm $1} & \text{0,$\pm $1,$\pm $2} & \text{0,$\pm $1} & \text{0,$\pm $1,$\pm $2} & \text{0,$\pm
   $1} & \text{0,$\pm $1,$\pm $2} & \text{0,$\pm $1} & \text{0,$\pm $1} & \text{0,$\pm $1,$\pm $2} \\ \hline \hline
 \text{phot/1000} & 22.2 & 31.8 & 10.2 & 14.8 & 7.3 & 10.8 & 9. & 39.7 & 57.4 \\
 \text{$f_G$} & 2.58 & 2.03 & 2.64 & 1.93 & 2.07 & 1.96 & 2.39 & 2.86 & 2.02 \\
\text{$\sigma_G$} & 1.67 & 1.71 & 1.27 & 1.40 & 1.40 & 1.27 & 0.70 & 1.14 & 1.32 \\
\text{$R_g$ (best fit)} & 18900 & 24000 & 14300 & 19600 & 13100 & 13800 & 7900 & 17000 & 24100 \\
\text{$R_g$ ($90\%$ l.l. from fit)} & 9100 & 10000 & 8000 & 8900 & 6200 & 6700 & 5300 & 10300 & 11600 \\
\text{$R_g$ ($90\%$ l.l. from $\Delta \chi^2$)} & 9800 & 11500 &\text{--} & \text{--} & \text{--} & \text{--} & \text{--} & 10400 & 12400 \\
\text{$R_g$ ($95\%$ l.l. from fit)} &  8300 & 9000 & 7400 & 8100 & 5600 & 6100 & 5000 & 9600 & 10600 \\
\text{$R_g$ ($95\%$ l.l. from $\Delta \chi^2$)} &  9500 & 10800 & \text{--} & \text{--} & \text{--} & \text{--} & \text{--} & 10000 & 11800 \\
\text{$L_{\rm Al}$} & 0.955 & 0.98 & 0.993 & 0.993 & 1.05 & 1.01 & 0.965 & 0.991 & 0.992 \\
\text{$\sigma_L$ from fit} & 0.034 & 0.027 & 0.035 & 0.03 & 0.06 & 0.051 & 0.055 & 0.023 & 0.019 \\
\text{$\sigma_L$ from $\Delta \chi^2$} &  0.04 &  +0.02,-0.04  & \text{--} & \text{--} & \text{--} & \text{--} & \text{--} & 0.03 & 0.02 \\
\text{$S_{\rm Al}$} &  0.995 & 0.974 & 0.959 & 0.971 & 0.921 & 0.911 & 0.931 & 0.969 & 0.969 \\
\text{$\sigma_S$} &  0.014 & 0.011 & 0.018 & 0.015 & 0.027 & 0.024 & 0.029 & 0.011 & 0.008 \\
\text{$\chi^2$} &  107.6 & 95.5 & 141.9 & 133.7 & 63.0 & 61.4 & 42.8 & 329.0 & 311.6 \\
\text{DOF} &  107 & 107 & 123 & 119 & 56 & 53 & 48 & 292 & 287 \\
\text{${\rm P}(\chi^2|{\rm DOF})$} & 0.53 & 0.22 & 0.88 & 0.83 & 0.76 & 0.80 & 0.31 & 0.93 & 0.85 \\
\end{tabular}}
  \caption{\footnotesize \textbf{Al with Variable Natural Line Width Fit Summary.}
All columns are Al-only. ($\sigma_G$, $\sigma_L$, $\sigma_S$) are $1\sigma$ errors for ($f_G$,  $L_{\rm Al}$, $S_{\rm Al}$). Effective resolving power values, $R_g$, use dispersion distance $D_{\rm Al18}$ for (a), (b), (h) \& (i), $D_{\rm Al14}$ for (c) \& (d), $D_{\rm Al10}$ for (e) \& (f), and $D_{\rm Al07}$ for (g). Best fit model comparisons with data are not shown for all cases because of similarity with other fits. Best fit models are compared with data in Figure \ref{AlFitFreeWd} for (a,b)
and \ref{AlFitComb2} for (i). $\Delta \chi^2$ analysis was not performed on cases (c)-(g), but probability contours are displayed for (a) and (b) in Fig.~\ref{AlFitcont1}, and (h) and (i) in Fig.~\ref{AlFitCombCont}.}
   \label{AlKfreeFit}
\end{table*}

\subsubsection{Ensemble Fitting of Mg-K and Al-K Orders.}

Due to the combination of lower counting rates and the target contamination requiring many additional fit parameters, the Mg data was of limited value in trying to estimate $R_g$. 
In our fit attempts of the 12$^{\rm th}$ order Mg data we found $f_G$ to be highly correlated with $S_{Mg}$.
When holding $S_{Mg}$ fixed at 0.93 (1$\sigma$ smaller than the nominal value based on the
uncertainties from Ref.~\cite{schweppe}), resulting in a worst-case estimate for $f_G$, we still obtained best fit values for $R_g \sim 8000-9000$.

When fitting Mg 12$^{\rm th}$ and Al 18$^{\rm th}$, 14$^{\rm th}$, and 10$^{\rm th}$ order data simultaneously we varied the Gaussian FWHM, $f_G$, and the Mg line width parameter $L_{\rm Mg}$ over specific grid values, while fitting most of the other parameters.
(We fixed the W-$M$ width ($L_W = 1.0$) and offset ($S_W = 1.0$) parameters and the Mg-$K\alpha_{1,2}$ separation parameter ($S_{Mg} = 0.93$) since they were ill-constrained.) Results for $R_g$ (best fit, 90\% and 95\% l.l.) were very similar to columns (h) and (i) in Table \ref{AlKfreeFit}, summarized in Table \ref{rsummary}.  This is not too surprising, given that the Mg data increased the total number of counts in the data set only by about 10\%, but the absence of a change in results for $R_g$ shows that the data from two different wavelengths are compatible (see Table \ref{rsummary}).

\subsubsection{Gratings X1 and X7}

Images were also collected with the Al target for gratings X1 (7$^{\rm th}$, 10$^{\rm th}$, 14$^{\rm th}$and 18$^{\rm th}$ order) and X7 (7$^{\rm th}$ and 10$^{\rm th}$ order) for comparison with X4. The total counts are much lower due to shorter integration times, constraining the effective resolving power lower limits to values smaller than results for X4.

\begin{figure}
 \begin{center}
 \includegraphics[width=3.5in]{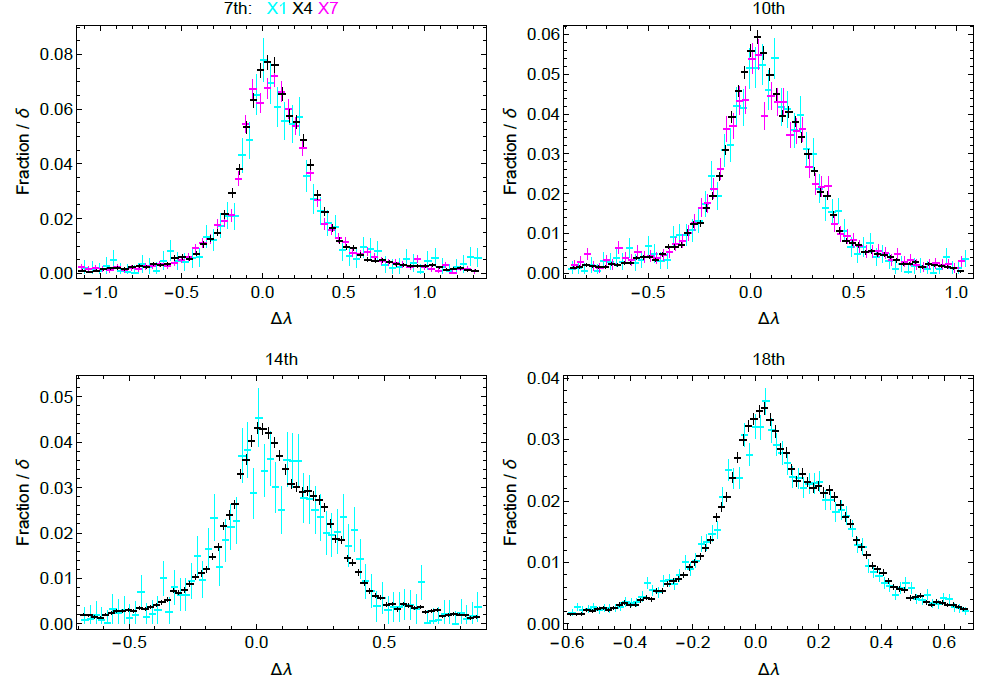}
 \end{center}
 \vspace{-0.2in}
\caption{\small  Al 7$^{\rm th}$, 10$^{\rm th}$, 14$^{\rm th}$ and 18$^{\rm th}$ line profile comparison among gratings. 7$^{\rm th}$ and 10$^{\rm th}$ order X4 (black) profiles are compared with X7 (magenta) and X1 (cyan) in the top panels. 14$^{\rm th}$ and 18$^{\rm th}$ order X4 profiles are compared with only X1 in the bottom panels. X7 efficiency was too low above 10$^{\rm th}$ order to justify measurements.}
 \label{AlOtherGratords}
\end{figure}

For comparison, identical orders from X1 and X7 are plotted on the same scale with X4 in Fig.~\ref{AlOtherGratords}. The profiles agree reasonably well. 

\begin{figure}
 \begin{center}
 \includegraphics[width=3.0in]{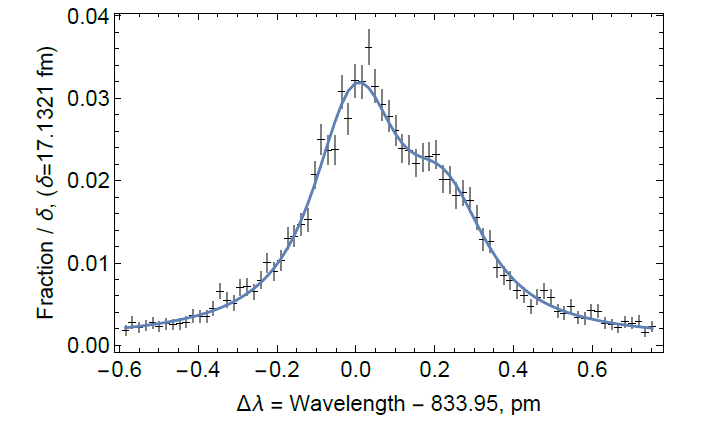}
 \end{center}
 \vspace{-0.2in}
\caption{\small  18$^{\rm th}$order Al best fit models for X1 grating using 5 CDB, compared with data. Line wavelengths and widths were fixed at 0.992 and 0.969 of their nominal values. 
}
 \label{AlFitX1Fixed}
\end{figure}

We analyzed the 18$^{\rm th}$ order X1 data in an abbreviated manner similar to X4.  First holding the Lorentzian and separation parameters constant at the values determined for X4 ($L_{\rm Al}=0.992$ and $S_{\rm Al}=0.969$), we performed a one dimensional $\Delta \chi^2$ analysis. The fit result is shown is shown in Fig.~\ref{AlFitX1Fixed}. Like X4, the result was consistent with $R_g=\infty$, with best fit $R_g=14000$. Due to larger statistical errors, the $90\%$ confidence lower limit was only 9000 compared with $>12000$ for X4 (see Table \ref{AlKfixedfit}).

For the remaining X1 and X7 cases we obtained $90\%$ c.l. lower ($R_g = 3300-9000$) and upper limits ($R_g = 10600 -\infty$) along with best fit values ($R_g = 6400 -\infty$) from 1-dimensional $\Delta \chi^2$ in the same manner.  

As expected, the $90\%$ lower limits are not as constraining as X4 results due to the lower count numbers.
However, all the resolving power confidence intervals overlap among the gratings for each order.  This suggests performance among gratings is consistent, even if we cannot obtain a lower limit $R_g>10000$ for X1 and X7 because of limited data. The results do indicate an effective resolving power $90\%$ c.l. lower limit of 9000 for X1, assuming the Al-K line parameters determined using X4.

\section{Discussion}
\label{discuss}
We conclude that the above results constrain the CAT grating effective resolving power above 11000 at $95\%$ confidence for X4.
The most constraining single order was the Al 18$^{\rm th}$ order using CDB:$0, \pm1, \pm2$ (Table \ref{AlKfreeFit}(b)) with $R_g$ ($95\%$ l.l. from $\Delta \chi^2$) $= 10800$.  We found that combining Al results from 14$^{\rm th}$ and 10$^{\rm th}$ order (Table \ref{AlKfreeFit}(i)) with those from 18$^{\rm th}$ order increased this limit to 11800. While Mg data was limited by spectral contamination and lower counting rates, we still found that combining the aforementioned Al orders with Mg 12$^{\rm th}$ order data yielded a comparable $R_g$ lower limit for both the CDB:$0, \pm1$ and CDB:$0, \pm1, \pm2$ cases.  Best fit and lower limit values are summarized in Table \ref{rsummary}. 

\begin{table}[h]
   \centering
{\tiny \begin{tabular}{lcccc}\hline 
\text{Fit ID} & \text{Table \ref{AlKfreeFit}(b)} & \text{Table \ref{AlKfreeFit}(i)} & \text{(c)} & \text{(d)}\\
\text{Order} & 18 & \text{Al-$\{$18,14,10$\}$} & \text{Al-$\{$18,14,10$\}$,Mg-12} & \text{Al-$\{$18,14,10$\}$,Mg-12} \\
\text{CDB} &\text{0,$\pm $1,$\pm $2} & \text{0,$\pm $1,$\pm $2} & \text{0,$\pm $1} & \text{0,$\pm
   $1,$\pm $2}  \\ \hline \hline
\text{$R_g$ (best fit)} & 24000 & 24100 & 16400 & 23200 \\
\text{$R_g$ ($90\%$ l.l. from fit)} & 10000 & 11600 & 10400 & 11700 \\
\text{$R_g$ ($90\%$ l.l. from $\Delta \chi^2$)} &11500 &12400  & 12700 & 12300\\
\text{$R_g$ ($95\%$ l.l. from fit)} &  9000 & 10600 & 9700 & 10700 \\
\text{$R_g$ ($95\%$ l.l. from $\Delta \chi^2$)} & 10800 &11800 & 11800 & 11700 \\
\end{tabular}}
  \caption{\footnotesize \textbf{Summary of best fit and lower limit effective resolving power results from the most sensitive fit cases.}
The $\Delta \chi^2$ values account for asymmetries in the confidence level contours and therefore, better reflect the true confidence levels.}
   \label{rsummary}
\end{table}

All of the other fits were consistent with these, in that lower sensitivity was the result of poorer statistics or lower dispersion.  There was typically less than a $10\%$ difference in the effective resolving powers derived from the central 3 vs.~central 5 CDBs.
Limits obtained from fit parameter error estimates were generally consistent with errors derived from $\Delta \chi^2$ analysis, with differences easily explainable by asymmetries in the error distribution.  All the cases were consistent with no grating contribution ($\infty$ effective resolving power) at the $90\%$ c.l. (upper limit) except for 7$^{\rm th}$ order Al.  Of course $R_g = \infty$ is only meaningful in the mathematical sense of our definition of $R_g = p/\Delta p$.  We neglect that the resolving power is always limited by the number of illuminated grating periods times the diffraction order.  Even for just 1 mm of illuminated width, $p = 200$ nm, and $m = 18$, this limit would mean $R < 90000$, which is practically indistinguishable from $R_g = \infty$ in our analysis.

Independent of x-ray data taken at the SLTF we have knowledge of certain properties of the three tested CAT gratings that should be considered when discussing the above results.

\subsection{Known Blaze Angle Variation.}  Three effects have an impact on what fraction of the area of a grating in our tests contributes efficiently to each diffraction peak. 

First, recent measurements using small-angle x-ray scattering (SAXS) have shown that our current DRIE tool produces deep etches with a small, but significant and systematic variation in etch angle across the surface of a 30 mm grating \cite{JungkiSPIE2018}.  This produces grating bars that are inclined relative to each other.  This means that, for any given orientation of the grating surface normal, grating bar sidewalls will be oriented at a range of angles to the incident x rays.  Thus some areas of the grating will be at the most favorable angle for blazing into $m^{\rm th}$ order, while others will not. For X4 we estimate from our efficiency models \cite{AO,SPIE2017} that the range of contributing angles is 0.65 (7$^{\rm th}$ order), 0.5 (10$^{\rm th}$ order), 0.3 (14$^{\rm th}$ order), and 0.25 degrees (18$^{\rm th}$ order).  The range becomes smaller due to the decreasing grating bar sidewall reflectivity with increasing $\alpha$. This translates into an area of 13 (7$^{\rm th}$ order), 10 (10$^{\rm th}$ order), 6 (14$^{\rm th}$ order) and 5 mm (18$^{\rm th}$ order) in length along the dispersion direction contributing to the given orders.  Recent testing with a more modern DRIE tool \cite{Rapier} has produced CAT grating geometry etches with significantly smaller etch angle variations \cite{Jungkiprep}.

Second, the gratings sit in a converging beam.  Over 30 mm in azimuth the incident angle varies by $\sim 0.19$ degrees.  In the mounted orientation (grating device layer facing the source) this effect runs in the opposite direction of the first effect, leading to a slight narrowing of the expected contributing areas.

Third, the gratings are not perfectly flat.  We observe a slight "dimpling" or "buckling" of the device layer within many L2 hexagons.  This effect amounts to a broadening of the blaze angle distribution across a hexagon, thus negating to some degree the impacts of the first two effects and increasing the range of areas contributing to a given diffraction peak.  Based on SAXS, white light interferometry \cite{SPIE2015}, and reflection measurements \cite{Jungki}, we estimate that this effect extends the range of contributing areas by 5 to 10 mm.

\subsection{Known Grating Period Variation.}  The gratings in this study were patterned using the optical interference of two mutually coherent spherical waves \cite{Ferrera}.  This creates a grating pattern with well-understood hyperbolic distortions and period distribution.  We can calculate this (non-Gaussian) period distribution from the geometric parameters of our interference lithography station and derive an upper limit for the resolving power.  If we consider the period distribution over a $(30 \times 4)$ mm$^2$ area in the center of the interference pattern we obtain $R_g = 26900$ as an upper limit.  If the interference pattern was off-center by 4 mm from the grating center (a reasonable possibility given our current procedures) the upper limit reduces to 18700.  Thus, even though the period distribution due to hyperbolic distortions is non-Gaussian, it is too narrow to have a noticeable impact on our effective resolving power modeling results.  In addition, period variations over the smaller areas that effectively contribute to the individual orders (see previous paragraph) are even smaller and can thus be safely ignored.  For future, much larger gratings there are alternative patterning techniques that do not suffer from hyperbolic distortions \cite{nanoruler}.

\subsection{Diffraction from the L1 Cross Support Mesh.}  The 5 $\mu$m period L1 cross support mesh is a periodic structure with a grating vector nominally perpendicular to the grating vector for the 200 nm-period CAT grating bars.  As a periodic structure it will also cause diffraction peaks, but in the direction along its own grating vector.  At each CAT grating diffraction order we would therefore expect first-order peaks from the L1 mesh at $\pm 114$ (Al-K) or $\pm 135$ pixels (Mg-K) in the cross-dispersion (y) direction.  Diffraction by the L1 mesh is expected to be weak due to the large L1 period (relative to the wavelength) and the small fraction of the period being taken up by the L1 bars.  In the  analysis presented here $\pm 1^{\rm st}$ order L1 peaks could potentially affect the outermost CDBs weakly if the L1 grating vector is not parallel to the y-direction (for example due to limited precision in the patterning steps during fabrication).  In other scenarios these peaks are more readily identifiable \cite{SPIE2017}.

\subsection{Estimating Period Distribution through Deconvolution.}  
\begin{figure}
 \begin{center}
 \includegraphics[width=2.5in]{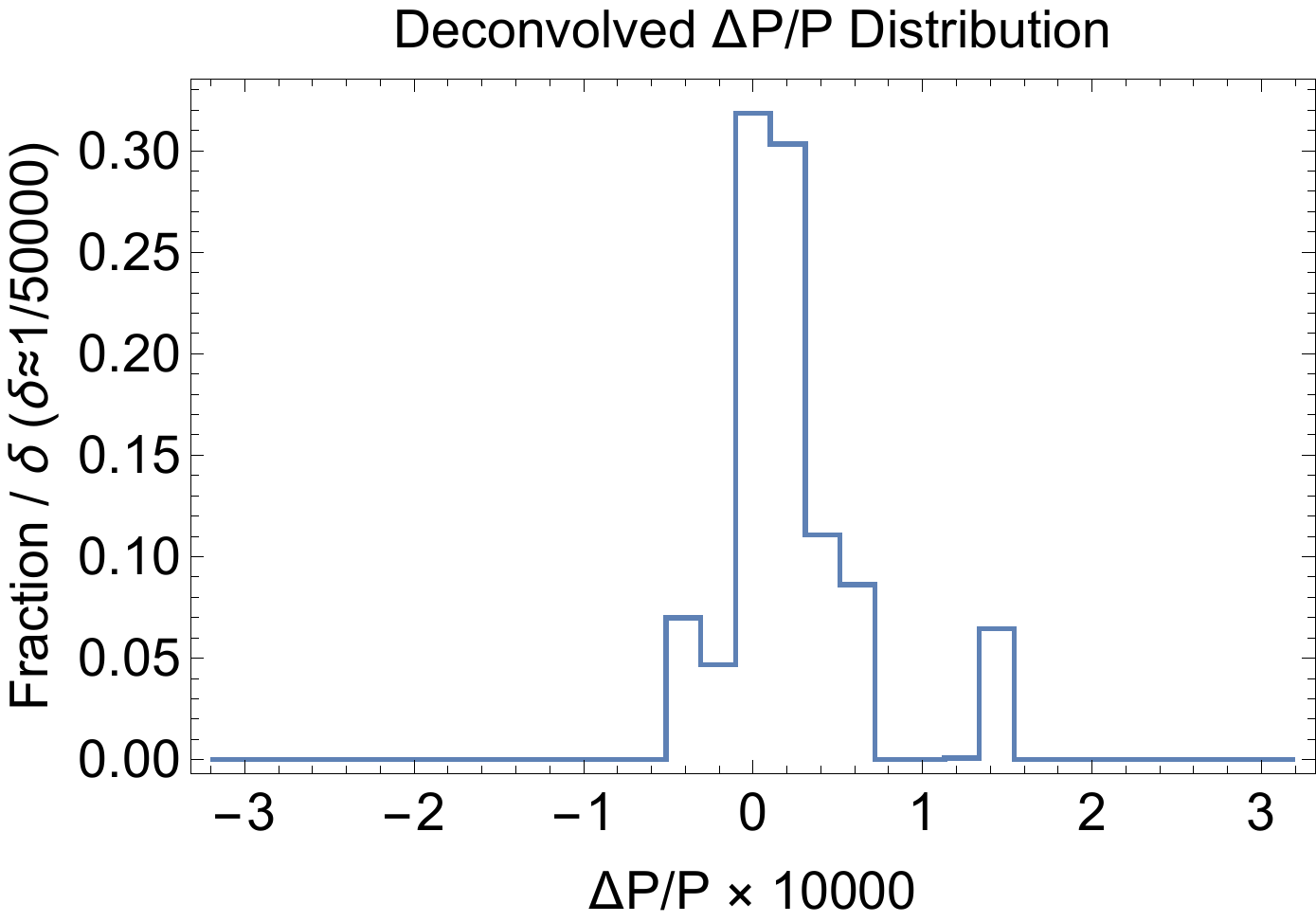}
  \end{center}
 \vspace{-0.2in}
\caption{\small  Distribution of period errors from deconvolution. Result is based on a Richardson-Lucy deconvolution of 18$^{\rm th}$ order Al data within the central 5 CDBs using the best fit line model convolved with zero order as the deconvolution kernel. The deconvolution was iterated until the variance between the re-convolved and measured profiles was minimized. 
}
 \label{rldeconvAl}
\end{figure}

To check if the grating response was significantly non-Gaussian within our measurement resolution, we performed a Richardson-Lucy deconvolution on the Al 18$^{\rm th}$ order measured line profiles, using the best fit line profile without Gaussian broadening. Under the assumption that the grating contribution is produced only by period variations, $\Delta p/p$, we can interpret the deconvolved profile as a distribution of periods relative to the nominal 200 nm. Fig. \ref{rldeconvAl} shows that the deconvolved grating contribution has most power constrained to two adjacent bins, which is consistent with an effective resolving power of 25000. While Gaussian FWHM errors were almost always large compared to the best fit value, we did find that the Al best fit values consistently fell between two and three pixels regardless of order. This might imply that the grating induced  broadening is not dominated by $\Delta p/p$ (which scales with dispersion distance). For example, if the L1 support structure is misaligned from the normal to the CAT grating bars, and cross-dispersion efficiency differs between zero order and other orders, then there could be a contribution which does not scale with dispersion and is not accounted for by the zero order. Since the test was performed with an essentially unfiltered Al spectrum, the cross-dispersion spectrum from zero order should contain all the continuum, and low level contaminant lines, including W-M and Mg-K, whereas the cross-dispersion at dispersed orders should contain only  Al-$K$ lines. So even if there is no cross-dispersion efficiency difference between zero order and other orders, there could still be a difference between the zero and dispersed order LRFs. Alternatively, there could be unknown systematic effects.

\subsection{Improved Al-K Doublet Parameter Uncertainties.} The Al-$K$ line width and separation parameters were best constrained by the combined Al 18$^{\rm th}$/14$^{\rm th}$/10$^{\rm th}$ orders using CDB:$0, \pm1, \pm2$ at $0.992\pm0.019$, and $0.969\pm0.008$, respectively. These correspond to $0.417\pm0.008\ {\rm eV}$ or $231.6\pm4.6\ {\rm fm}$ for the line width, and $0.400\pm0.003\ {\rm eV}$  or $224.4\pm1.8\ {\rm fm}$ for the line separation. The line width includes uncertainty due to the zero order width uncertainty of 1.2 fm or 0.002 eV.  This separation uncertainty is a factor of 3 smaller than the value quoted in  Ref.~\cite{schweppe}, and the line width uncertainty is a factor of 2.5 smaller than the value quoted in Ref.~\cite{campbell}.  

We made the following assumptions in the interpretation of the fit results:

1. We assumed a Gaussian response to account for the dispersed grating contribution.
2. The natural line shape is a pure Lorentzian function.
3. The convolution of the natural line shapes with the zero order response function represents the predicted response from a grating with $R_g=\infty$.
4. Broadening in the dispersed profile is assumed to be a measure of $\Delta p/p$ errors in the gratings.

The strongest and only result contradicting the $\Delta p/p$ assumption is 7$^{\rm th}$ order Al, which indicates a $> 3 \sigma$ detection of a 2.4 pixel FWHM Gaussian contribution in the dispersed profile. Scaling, this would imply a 6.2 pixel FWHM at 18$^{\rm th}$ order, or $R_g\approx8000$, which is excluded at $>99\%$ c.l.  On the other hand, effective resolving powers of $\{11000,20000\}$ at 18$^{\rm th}$ order, would suggest Gaussian FWHMs of $\{1.7, 1.0\}$ pixels which are only $\{$1.0$\sigma$, 2.0$\sigma\}$ from the best fit value for 7$^{\rm th}$ order. So even the 7$^{\rm th}$ order measurement is reasonably consistent with the other results.

\section{Summary and Conclusions}

We assembled a breadboard prototype for a CAT grating-based x-ray spectrometer, consisting of a grazing incidence Wolter-I mirror pair (TDM) and 30 mm-wide CAT gratings.  The TDM was illuminated with x rays from an electron impact source with Al and Mg targets at a distance of 92.17 m.  The convolution of slit-limited source, TDM PSF, and CCD pixelation provided an anisotropic focal spot with 0.9 arcsec FWHM in the horizontal direction.  

Zeroth order line profiles (with a grating moved into the beam) only showed weakly increased scatter compared to the unobstructed line profile, demonstrating minimal changes in the PSF due to the gratings.  We developed an empirical line response function model for zeroth order for convolution with the spectrum of the characteristic K$_{\alpha}$ doublets for Al and Mg.  Measured values from the literature were used as reference parameters for the doublets. The resulting modeled response served as prediction for the measured diffracted line profile.  We assumed any additional broadening to be due to grating period variations and modeled this effect as an additional convolution with a Gaussian with FWHM of $f_G$.  Gratings with period variation width $f_G$ would limit resolving power of the measurement to $R_g = D_{mi}/f_G$.  Fits to our data up to 18$^{\rm th}$ order for Al-K consistently produced $R_g > 11000$ at 95\% confidence for grating X4, and best fit values in the range of $R_g \sim 16000 - 24000$.  

For 18$^{\rm th}$ order Al the maximum resolving power $R_{XGS}$ for the spectrometer could be $\sim 24000$, only utilizing the central three CDBs, and assuming no broadening due to the grating. Taking the deduced grating contribution into account from column (c) in Table \ref{rsummary}, for example, we arrive at an experimentally demonstrated 90\% confidence level lower limit for the resolving power of $R_{XGS} \sim 11200$ for the presented breadboard grating spectrometer. 

Fits to data from gratings X1 and X7 were consistent with the results from X4, but produced lower constraints on $R_g$ due to shorter integration times and therefore lower counting statistics. Within measurement uncertainty the three independently fabricated gratings provide the same resolving power in a given diffraction order, lending confidence to the repeatability of the fabrication process.

The best constraints often resulted in values for the relative Al-K linewidth and separation parameters to be slightly less than one. The 1$\sigma$ uncertainties usually include the value one for the linewidth parameter, but slightly less so for the separation parameter.  The smaller uncertainties, compared to literature values, indicate that the spectrometer setup in this work provides superior resolving power compared to a soft x-ray double crystal spectrometer.  This performance might inspire new high-resolution spectrometer designs for laboratory-based plasma and astrophysics studies.

Our results demonstrate that CAT gratings are compatible with XGS designs with resolving power $R_{XGS} > 10000$.  The relaxed alignment and figure requirements in the transmission geometry tolerate many micrometers of non-flatness in the grating membranes and many arcminutes misalignment for most rotational degrees of freedom, which reduces constraints on fabrication, alignment and mounting.  CAT gratings have passed environmental testing for launch vibrations and temperature cycling under vacuum without performance degradation, and grating-to-grating roll alignment to within 5 arcmin has been demonstrated \cite{SPIE2017,SPIE2018}.  Significant gains in diffraction efficiency are possible from CAT gratings with deeper ($\sim 6$ $\mu$m) grating bars \cite{SPIE2016}. 
CAT gratings are therefore promising diffraction gratings for space-based soft x-ray spectrometers with high resolving power and large collecting area, and they can easily meet requirements for the Arcus Explorer mission ($R_{XGS} > 2500$) currently undergoing a NASA Phase A study \cite{Arcus,SPIE2018}.

\paragraph{Funding.} 
National Aeronautics and Space Administration (NASA) (NNX15AC43G).

\paragraph{Acknowledgments.}  
We like to express our gratitude to W.~Zhang, R.~McClelland, K.-W.~Chan, J.~Niemeyer, and M.~Schofield from the Next-Generation X-ray Optics group at NASA Goddard Space Flight Center for supplying the TDM, to S.~O'Dell (NASA Marshall Space Flight Center) for helpful discussions, to R.~Bhatia (Veeco-CNT) for ALD coating of the CAT gratings, and to J.~Song (MIT) for SAXS measurements.

{}

\end{document}